\documentclass[useAMS,usenatbib,usegraphicx]{mn2e}

\usepackage{times} 

\title[Testing model predictions of CDM with the SDSS]{Testing 
model predictions of the cold dark matter cosmology for the sizes, colours,
morphologies and luminosities of galaxies with the SDSS}
\author[J. E. Gonz\'{a}lez et al.]{Juan
E. Gonz\'{a}lez$^{1}$\thanks{E-mail: j.e.gonzalez@durham.ac.uk},
C. G. Lacey$^{1}$, C. M. Baugh$^{1}$, C. S. Frenk$^{1}$,
A. J. Benson$^2$\\ 
$^{1}$Institute for Computational Cosmology,
Department of Physics, Durham University, South Road, Durham DH1 3LE,
UK\\ 
$^2$Mail Code 130-33, California Institute of Technology,
Pasadena, CA 91125, USA}
\begin{document}

\date{}

\pagerange{\pageref{firstpage}--\pageref{lastpage}} \pubyear{2008}

\maketitle
 
\label{firstpage}

\begin{abstract}
The huge size and uniformity of the Sloan Digital Sky Survey makes
possible an exacting test of current models of galaxy formation. We
compare the predictions of the {\tt GALFORM} semi-analytical galaxy
formation model for the luminosities, morphologies, colours and
scale-lengths of local galaxies. {\tt GALFORM} models the luminosity
and size of the disk and bulge components of a galaxy, and so we can
compute quantities which can be compared directly with SDSS
observations, such as the Petrosian magnitude and the S\'{e}rsic index.
We test the predictions of two published models set in the cold dark
matter cosmology: the \citet{Bau05} model, which assumes a top-heavy
initial mass function (IMF) in starbursts and superwind feedback, and
the \citet{Bow06} model, which uses AGN feedback and a standard IMF.
The Bower et al model better reproduces the overall shape of the
luminosity function, the morphology-luminosity relation and the colour
bimodality observed in the SDSS data, but gives a poor match to the
size-luminosity relation. The \citeauthor{Bau05} model successfully
predicts the size-luminosity relation for late-type galaxies. Both
models fail to reproduce the sizes of bright early-type
galaxies. These problems highlight the need to understand better both
the role of feedback processes in determining galaxy sizes, in
particular the treatment of the angular momentum of gas reheated by
supernovae, and the sizes of the stellar spheroids formed by galaxy
mergers and disk instabilities.
\end{abstract}

\begin{keywords}
galaxies: evolution --- galaxies: formation --- methods: numerical
\end{keywords}

\section{INTRODUCTION}

There is a growing weight of evidence in support of the hierarchical
paradigm for structure formation \citep{Spr06}.  The principal process
responsible for the growth of structure, gravitational instability,
has been modelled extensively using large numerical simulations
\citep[e.g.][]{Spr05,Spr08}. The cold dark matter model gives an
impressive fit to measurements of temperature fluctuations in the
cosmic microwave background radiation \citep{Hinshaw08}. When combined
with other data, such as measurements of local large scale structure
in the galaxy distribution or the Hubble diagram of type Ia
supernovae, there is a dramatic shrinkage in the available range of
cosmological parameter space \citep{Per02,Ariel06,Kom08}. The
spectacular progress made in constraining the background cosmology has
allowed the focus to shift to trying to understand the evolution of
the baryonic component of the Universe \citep{Bau06}.

Over the same period of time there has been a tremendous increase in
the quantity and range of observational data on galaxies at different
redshifts. Observations at high redshift have uncovered populations of
massive, actively star forming galaxies which were already in place
when the Universe was a small fraction of its current age
\citep[e.g.][]{Sm97,St99,Blain02,Gia02}. Huge surveys of the local
universe made possible by advances in multifibre spectrographs allow
the galaxy distribution to be dissected in numerous ways
\citep[e.g.][]{Colless03,Adel08}. It is now possible to make robust
measurements of the distribution of various intrinsic galaxy
properties, such as luminosity, colour, morphology and size. The
trends uncovered in these local surveys are influenced by a wide range
of physical effects, such as star formation, supernova and AGN
feedback, the cooling of gas and galaxy mergers, and hence provide as
strict a test of galaxy formation models as that posed by the high
redshift data.

In order to test current ideas about galaxy formation against
observations, well developed theoretical tools are needed which can
follow many complex processes concurrently. Most importantly, it is
vital that the model predictions are produced in a form which can be
compared as directly as possible with observations. Gas dynamic
simulations typically follow galaxy formation in great detail for an
individual dark matter halo \citep[e.g.][]{Gov04,Takashi05,Gov07} or
within some small volume \citep[e.g.][]{Nag04,Sca06,Croft08}. In
general, only limited output is available which can be compared
directly with observations, such as, for example, estimates of galaxy
luminosity.  Currently the closest contact with observations is made
by semi-analytical models of galaxy formation (see \citet{Bau06} for a
review). In their most sophisticated form, these models can make
predictions for the luminosity, colour, scale-length and morphology of
galaxies in a wide range of environments
\citep[e.g.][]{Bow06,Cro06,Catt06,Monaco07,Claudia08}. These models
necessarily have to treat the baryonic physics in a somewhat more
idealized way than is done in the gas dynamic simulations. Physical
processes are described using rules, some of which contain parameters
whose values are set by comparing the model predictions with selected
observations. As the range of data the model is compared against
increases, the parameter space open to the models reduces. For
example, the strength of supernova feedback (see Section 2) has an
impact not only on the number of galaxies in the faint end of the
luminosity function, but also on the size of galactic disks and even
the morphological mix of galaxies.

Two key properties of semi-analytical models are their computational
speed and modular nature. The impact of different processes on the
nature of the galaxy population can be rapidly assessed by running
models with different parameter choices. This makes the models ideal
tools with which to interpret observational data. Any discrepancy
uncovered between the model predictions and observations can help to
identify physical ingredients which may either require better
modelling or which may be missing altogether from the calculation. One
clear example of how observations drive the development of the models
is given by the recent efforts to reproduce the location and sharpness
of the break in the galaxy luminosity function.  With the current best
fitting cosmological parameters, galaxy formation models struggled to
avoid producing too many bright galaxies \citep{Ben03}. One solution
to this problem was suggested by observations showing the apparent
absence of cooling flows at the centres of rich clusters
(e.g. \citealt{Pet03}) which motivated the idea of taking into account
the energy released by active galactic nuclei (AGN). This acts as a
feedback process that heats the gas in massive haloes. The
incorporation of this feedback mechanism suppresses the formation of
galaxies in massive haloes, such that the right number of bright
galaxies can be produced, and, furthermore, these galaxies have red
colours to match those observed \citep{Bow06,Cro06,Claudia08}.

In this paper we test two published galaxy formation models run using
the {\tt GALFORM} semi-analytical model against statistics measured
from the Sloan Digital Sky Survey. The \citet{Bau05} model (hereafter
Baugh2005) invokes a ``superwind'' channel for supernova feedback,
which ejects gas from low and intermediate mass haloes. This model
assumes that stars are produced with a normal initial mass function
(IMF) in quiescent disks but with a top-heavy IMF during merger-driven
starbursts. The Baugh2005 model is able to reproduce the counts and
redshift distribution of sub-millimetre selected galaxies, along with
the abundance of Lyman-break galaxies and Lyman-$\alpha$ emitters
\citep{LeD05,LeD06,Orsi08}. The \citet{Bow06} model (hereafter
referred to as Bower2006) incorporates AGN feedback, with the energy
released by the accretion of mass onto the central supermassive black
hole in halos with quasistatic hot gas atmospheres being responsible
for stifling the cooling rate. The Bower2006 model gives a good match
to the evolution of the K-band luminosity function and the inferred
stellar mass function.

The paper is organized as follows. In Section 2, we outline the galaxy
formation model {\tt GALFORM} tested in the paper.  In Section 3 we
describe how some additional galaxy properties are computed from the
model output; these properties are needed to compare the model
predictions directly with observations of SDSS galaxies.  Section 4
contains the comparisons between model predictions and SDSS data for
the luminosity function, the distribution of morphological types, the
colour distribution and the size distribution. In this section we also
show the impact on the predictions of changing the strengths of
various processes in the model. In Section 5 we present our
conclusions.  The appendices discuss how certain photometric
properties of galaxies have changed between SDSS data releases and
compare different indicators of galaxy type. Finally, we note that
magnitudes are quoted on the AB system assuming a Hubble parameter of
$h=H_{0}/100$ km s$^{-1}$ Mpc$^{-1}$; the cosmological parameters
adopted depend on the choice of semi-analytic model as explained in
Section 2.

\section[]{GALAXY FORMATION MODEL}
We use the Durham semi-analytical galaxy formation model, {\tt
GALFORM}, introduced by \citet{Col00} and developed in a series of
subsequent papers \citep{Ben03,Bau05,Bow06}. The model tracks the
evolution of baryons in the cosmological setting of a cold dark matter
universe. The physical processes modelled include: i) the hierarchical
assembly of dark matter haloes; ii) the shock heating and
virialization of gas inside the gravitational potential wells of dark
matter haloes; iii) the radiative cooling of the gas to form a
galactic disk; iv) star formation in the cool gas; v) the heating and
expulsion of cold gas through feedback processes such as stellar winds
and supernovae; vi) chemical evolution of gas and stars; vii) mergers
between galaxies within a common dark halo as a result of dynamical
friction; viii) the evolution of the stellar populations using
population synthesis models; ix) the extinction and reprocessing of
starlight by dust.  In this section we first give a comparison of the
main features of the Baugh2005 and Bower2006 models, introducing the
recipes used to model phenomena which are varied in Section 4. Similar
discussions of these models can be found in \citet{Al07}, \citet{Al08}
and \citet{Violeta08}. In the second part of this section we review
the model used to compute galaxy sizes, which was originally devised
by \citet{Col00} and tested by these authors for galactic disks and by
\citet{Al07} for spheroids.

\subsection[]{A comparison of the Baugh2005 and Bower2006 models}
The Baugh2005 and Bower2006 models represent alternative models of
galaxy formation. The parameters which specify the models were set by
the requirement that their predictions should reproduce a subset of
the available observations of local galaxies together with certain
observations of high redshift galaxies. Different solutions were found
due to the use of different physical ingredients, as set out below,
and because different emphasis was placed on reproducing particular
observations. We refer the reader to the original papers for a full
description of each model; the Baugh2005 model is also described in
detail in \citet{Lacey08}.

We now summarize the main differences between the two models. 

\begin{itemize}
\item {\it Cosmology.}
The Baugh2005 model adopts a $\Lambda$CDM cosmology with a present-day 
matter density parameter, $\Omega_{0}$=0.3, a cosmological constant, 
$\Omega_{\lambda}$=0.7, a baryon density, $\Omega_{b}=0.04$ and a 
power spectrum normalization given by $\sigma_{8}=0.93$. The Bower2006 
model uses the cosmological model assumed in the Millennium simulation 
(\citealt{Spr05}), where $\Omega_{0}$=0.25, $\Omega_{\lambda}$=0.75, 
$\Omega_{b}=0.045$ and $\sigma_{8}=0.9$, which are in somewhat better 
agreement with the constraints from the anisotropies in the cosmic 
microwave background and galaxy clustering on large scales 
(e.g. \citealt{Ariel06}).

\item {\it Halo merger trees.}  The Baugh2005 model uses a Monte-Carlo
technique to generate merger histories for dark matter haloes, which
is based on the extended Press-Schechter theory
\citep{Lacey93,Col00}. The formation and evolution of a representative
sample of dark matter haloes is followed. In the Bower2006 model, the
merger histories are extracted from the Millennium simulation (see
\citealt{Har06} for a description of the trees). The mass resolution
of the simulation trees is $1.72 \times 10^{10} h^{-1} M_{\odot}$,
which is a factor of three worse than that used in the Monte-Carlo
trees. By comparing the output of models using Monte-Carlo and N-body
merger trees, \citet{Helly03} found very similar predictions for
bright galaxies, with differences only becoming apparent below some
faint magnitude, the value of which depends on the mass resolution of
the N-body trees.  The resolution of the Millennium simulation yields
a robust prediction for the luminosity function to around three
magnitudes fainter than $L_{\star}$, which is more than adequate for
the comparisons presented in this paper.

\item{\it Quiescent star formation timescale.}  The quiescent star
formation rate in disks, $\psi$, is given by $\psi = M_{\rm
gas}/\tau_{\ast}$, where $M_{\rm gas}$ is the mass of cold gas and the
time-scale, $\tau_{\ast}$, is parameterized differently in the two
models. In both cases, the star formation timescale is allowed to
depend upon some power of the circular velocity of the disk and is
multiplied by an efficiency factor. In the Baugh2005 model, the
efficiency factor is assumed to be independent of redshift, whereas in
the case of the Bower2006 model, this factor scales with the dynamical
time of the galaxy \textbf{[$\tau_{\rm dyn}$]}, measured at the half-mass radius of the
disk. Since the typical dynamical time gets shorter with increasing
redshift, the star formation timescale in the Bower2006 model is
shorter at high redshift than it would be in the equivalent disk in
the Baugh2005 model. This has implications for the amount of star
formation in merger-triggered starbursts (or following a disk becoming
dynamically unstable in the Bower2006 model - see later).  In the
Bower2006 model, disks at high redshift tend to be gas poor, with the
gas being turned into stars on a short timescale after cooling,
whereas in the Baugh2005 model, high redshift disks are gas rich.

\item {\it Initial mass function (IMF) for star formation.}  The
Bower2006 model uses the \citet{Ken83} IMF, consistent with deductions
from the solar neighbourhood, in all modes of star formation. The
Baugh2005 model also adopts this IMF in quiescent star formation in
galactic disks.  However, in starbursts triggered by galaxy mergers, a
top-heavy IMF is assumed. The yield of metals and the fraction of gas
recycled per unit mass of stars formed are chosen to be consistent
with the form of the IMF.

\item {\it Supernova (SN) feedback.}
With each episode of star formation, a mass of cold gas is reheated 
and ejected from the disk by supernova explosions at a rate given by:
\begin{equation}
\dot{M}_{\rm eject}=(V_{\rm hot}/V_{\rm disk})^{\alpha_{\rm hot}} \psi, 
\label{eqsnfeedback}
\end{equation}
where $V_{\rm disk}$ is the velocity at the disk half-mass radius, and
$V_{\rm hot}$ and $\alpha_{\rm hot}$ are parameters.  The SN feedback
is stronger in the Bower2006 model ($V_{\rm hot}=485 {\rm km s}^{-1}$
and $\alpha_{\rm hot} = 3.2$ compared with $V_{\rm hot}=300 {\rm km
s}^{-1}$ and $\alpha_{\rm hot} =2$ in the Baugh2005 model).

\item {\it AGN vs superwind feedback.}  Perhaps the most significant
difference between models is the manner in which the formation of very
massive galaxies is suppressed. In the Baugh2005 model, an additional
channel or fate for gas heated by supernovae is invoked, called
superwind feedback. In addition to the standard SNe feedback model
described in the previous bullet point, some gas is assumed to be
expelled completely from the halo due to heating by supernovae.  The
amount of mass ejected is taken to be a multiple of the star formation
rate, multiplied by a function of the circular velocity of the
halo. The superwind is most effective in removing gas from low
circular velocity haloes, with the mass of gas ejected falling with
increasing circular velocity. The gas expelled in the superwind is not
allowed to recool, even in more massive haloes at later times in the
merger history.  This has the effect of reducing the cooling rate in
massive haloes since these haloes have less than the universal
fraction of baryons. Such winds have been observed in massive
galaxies, with the inferred mass ejection rates found to be comparable
to the star formation rate (e.g. \citealt{Heckman90,Pett01,Wil05}).  In the
Bower2006 model an AGN feedback model is implemented which regulates
the cooling rate, effectively switching off the supply of cold gas for
star formation in quasi-static hot gas haloes.  These are haloes in
which the cooling time of the gas exceeds the free-fall time within
the halo. The cooling flow is quenched by the energy injected into the
hot halo by the central AGN. The growth of the black hole is followed
using the model described by \citet{Rowena07}

\item {\it Disk instabilities.}  In the Baugh2005 model the only
process that leads to the formation of bulge stars is a galaxy merger.
In the Bower2006 model, strongly self-gravitating disks are considered
to be unstable to small perturbations, such as encounters with minor
satellites or dark matter substructures. Such events can lead to the
formation of a bar and eventually the disk is transformed into a
bulge. The onset of instability is governed by the ratio
\begin{equation}
\epsilon =  \frac{V_{\rm disk}}{(GM_{\rm disk}/r_{\rm disk})^{1/2}}.
\label{eqinstab}
\end{equation}
Disks for which $\epsilon < \epsilon_{\rm disk}$, are considered to be
unstable; in Bower2006, the threshold for unstable disks is set at
$\epsilon_{\rm disk} = 0.8$. Any cold gas present when the disk
becomes unstable is assumed to participate in a starburst. As with
starbursts triggered by galaxy mergers, a small fraction of the gas
involved in the burst is accreted onto the central black hole. This is
an important channel for the growth of low and intermediate mass black
holes in the Bower2006 model.

\item {\it Treatment of reheated cold gas.}  The fate of gas reheated
by supernovae is different in the two models.  In the Baugh2005 model,
as discussed above, there are two possible fates for the gas heated by
supernovae: ejection from the disk to be reincorporated into the hot
halo and ejection from the halo altogether, with no possibility of
recooling at a later time. In the case of the first of these channels,
the gas is added back into the hot halo when a new halo forms. This
happens when the original halo has doubled in mass since its formation
time. In the Bower2006 model, this timescale is instead taken to be
some multiple of the dynamical time of the halo. Thus, gas can be
reheated by supernovae, be added back into hot halo and cool again on
a shorter timescale in the Bower2006 model than in the Baugh2005
model.
  
\end{itemize}

\subsection{Galaxy scale-lengths}
For completeness, We now review the calculation of the sizes of the
disk and bulge components of galaxies used in {\tt GALFORM}, to
complement the discussion of the size predictions presented in Section
4.  For a more detailed description we direct the reader to Cole et~al
(2000).  In the following subsections, we outline how the
scale-lengths of the disk and bulge are calculated. The scale-lengths
are determined by solving for the dynamical equilibrium of the disk,
bulge and dark matter together.  Dark matter haloes are assumed to
have a \citet{NFW97} density profile. The hot halo has a modified
isothermal density profile with a core.

\subsubsection{Disks}

The size of a galactic disk is determined by the conservation of the
angular momentum of the gas cooling from the halo and the application
of centrifugal equilibrium.  The disk is assumed to have an
exponential surface density profile with half-mass radius $r_{\rm
disk}$. The half-mass radius of the disk is related to the exponential
scale-length, $h_{\rm D}$ by $r_{\rm disk} = 1.68 h_{\rm D}$.

The angular momentum of the dark matter halo is quantified by a
dimensionless spin parameter which is drawn from a log-normal
distribution, as suggested by measurements from high resolution N-body
simulations \citep{Col96}. This scheme is used in both models; the
spin parameter of low-mass haloes can not be measured reliably from
the Millennium simulation \citep{Bett07}. Each newly formed halo in a
halo merger tree is assigned a new spin parameter drawn from the
distribution at random, independently of the previous value of the
spin parameter.

As the halo collapses, the associated gas will shock and so be heated
to approximately the virial temperature of the halo. Thereafter it
will begin to cool via atomic processes. As gas cools it loses
pressure support and flows to the center of the halo, where it is
assumed to settle into a rotationally supported disk. The model
assumes that the specific angular momentum of the cooling gas depends
on the radius from which it originated, as described in \citet{Col00}.
The scale-length of the disk is dependent on the angular
momentum of the halo gas which cools.

\subsubsection{Bulges}

Spheroids result from galaxy mergers, and in the case of the Bower2006
model also from dynamical instabilities of disks. The size of the
spheroid produced by these events is calculated by considering virial
equilibrium and energy conservation. We assume that the projected
density profile is well described by a de Vaucouleurs $r^{1/4}$ law
(e.g. Binney \& Tremaine 1987). The effective radius, $r_{\rm e}$, of
the $r^{1/4}$ law, i.e. the radius that contains half the mass in
projection, is related to the half-mass radius in three dimensions,
$r_{\rm bulge}$, by $r_{\rm bulge}=1.35 r_{\rm e}$.

When dark matter haloes merge, the galaxy in the most massive halo is
assumed to become the central galaxy in the new halo while the other
galaxies become satellites. The orbits of the satellites gradually
decay as energy and angular momentum are lost via dynamical
friction. Eventually a satellite will merge with the central galaxy if
the timescale for the orbit to decay is shorter than the halo
lifetime.

Two types of merger are distinguished, major mergers and minor
mergers, according to the ratio of the mass of the smaller galaxy to
the larger galaxy $M_2/M_1$. If the ratio is $M_2/M_1\geq f_{\rm
ellip}$, then a major merger is assumed to have taken place. Both
stellar disks are transformed into spheroid stars and added to any
pre-existing bulge. Any cold gas present takes part in a starburst. If
the ratio is $M_{\rm 2}/M_{\rm 1}\leq f_{\rm ellip}$ then a minor
merger is assumed and the stars of the smaller galaxy are added to the
bulge of the central galaxy, which keeps its stellar disk.  If $f_{\rm
ellip} > M_{\rm 2}/M_{\rm 1}\geq f_{\rm burst}$ and the central galaxy
is gas rich, then the minor merger may also be accompanied by a
burst. The parameter $f_{\rm ellip}=0.3$ in both models; $f_{\rm
burst}=0.05$ in Baugh2005 and $f_{\rm burst} = 0.1$ in
Bower2006. Disks are considered gas rich if 10\% of the total disk
mass in stars and cold gas is in the form of cold gas in Bower2006;
this threshold is set higher, 75\%, in Baugh2005.

In a merger, the two galaxies are assumed to spiral together until
their separation equals the sum of their half-mass radii, which is the
moment when they are considered to have merged. We estimate the radius
of the merger remnant using energy conservation. Assuming virial
equilibrium, the total internal energy $E_{\rm int}$ of each galaxy
(both for the merging components and the remnant of the merger) is
related to its gravitational self-binding energy $U_{\rm int}$ by
$E_{\rm int} = -\frac{1}{2} U_{\rm int}$, and so can be written as
\begin{equation}
E_{\rm int}=-\frac{\bar{c}}{2} \frac{GM^2}{r},
\label{Ebind}
\end{equation}
where $M$ and $r$ are the mass and half-mass radius respectively and
$\bar{c}$ is a form factor which depends on the distribution of mass
in the galaxy. For a de Vaucouleurs profile, $\bar{c}=0.45$, while for
an exponential disk, $\bar{c}=0.49$. For simplicity, in the model we
assume $\bar{c}=0.5$ for all galaxies.
 
The orbital energy of a pair of galaxies at the moment of merging is given by
\begin{equation}
E_{\rm orbit}=-\frac{f_{\rm orbit}}{2}\frac{GM_{1}M_{2}}{r_{1}+r_{2}},
\label{Eorbit}
\end{equation}
where $M_{1}$ and $M_{2}$ are the masses of the merging galaxies, $r_{1}$ 
and $r_{2}$ are their half-mass radii, and $f_{\rm orbit}$ is a parameter 
which depends on the orbital parameters of the galaxy pair.
A fiducial value of $f_{\rm orbit}=1$ is adopted which 
corresponds to two point mass galaxies in a circular orbit with 
separation $r_{1}+r_{2}$. The galaxy masses $M_{1}$ and $M_{2}$ include 
the total stellar and cold gas masses and also some part of the dark
matter halo. We assume that the mass of dark matter which participates in
the galaxy merger in this way is a multiple $f_{\rm dark}$ of dark
halo mass $M_{i,\rm dark}$ within the galaxy half-mass radius. We
adopt a fiducial value $f_{\rm dark}=2$. We thus have:
\begin{equation}
M_{i}=M_{i,\rm stellar+gas}+f_{\rm dark} M_{i,\rm dark}.
\label{DMcontrib}
\end{equation}
Later on we will investigate the effect of varying $f_{\rm orbit}$ and
the dark matter mass contribution $f_{\rm dark}$.

Assuming that each merging galaxy is in virial equilibrium, then their
total energy equals one half of their internal energy. The
conservation of energy means that:
\begin{equation}
E_{\rm int,new}=E_{\rm int,1}+E_{\rm int,2}+ E_{\rm orbit},
\label{Econservation}
\end{equation}
and replacing $E_{\rm int}$ with Eq.~\ref{Ebind} and $E_{\rm orbit}$
with Eq.~\ref{Eorbit} leads to
\begin{equation}
\frac{(M_{1}+M_{2})^2}{r_{\rm new}}=\frac{M_{1}^2}{r_{1}}+\frac{M_{2}^2}{r_{2}}+\frac{f_{\rm orbit}}{\bar {c}}\frac{M_{1}M_{2}}{r_{1}+r_{2}},
\label{rnew}
\end{equation}
where $r_{\rm new}$ is the half-mass radius of the remnant immediately
after the merger. These equations lead to the result that, in a merger
of two identical galaxies, the half-mass radius of the new galaxy
increases by a factor of $4/3$ which agrees reasonably well with the
factor of 1.42 found in simulations of equal mass galaxy mergers by
Barnes (1992).

In the case of a bulge produced by an unstable disk (which is only
considered in the Bower2006 model), the considerations that lead to
the remnant size are similar to those applied to a bulge produced by
mergers, leading to:
\begin{eqnarray}
\frac{\bar{c}_{\rm B}(M_{\rm disk}+M_{\rm bulge})^2}{r_{\rm new}} & = & \frac{\bar {c}_{\rm B}M_{\rm bulge}^2}{r_{\rm bulge}}+\frac{\bar {c}_{\rm D}M_{\rm disk}^2}{r_{\rm disk}}\\
& + & {f_{\rm int}}\frac{M_{1}M_{2}}{r_{1}+r_{2}}, \nonumber
\label{rnewinstab}
\end{eqnarray}
where $M_{\rm bulge}$, ${r_{\rm bulge}}$ and $M_{\rm disk}$, ${r_{\rm
disk}}$ refer to the masses and half-mass radii of the bulge (if any)
and disk respectively. As mentioned above, the form factors $\bar {c}_{\rm
B}=0.49$ and $\bar{c}_{\rm D}=0.45$ for a bulge and disk respectively. The
last term in Eq.~\ref{rnewinstab} represents the gravitational
interaction energy of the disk and bulge, which can be approximated
for a range of $r_{\rm bulge}/r_{\rm disk}$ with $f_{\rm int}=2$.

\section[]{DERIVED GALAXY PROPERTIES}

\begin{figure}
\includegraphics[width=8.5cm]{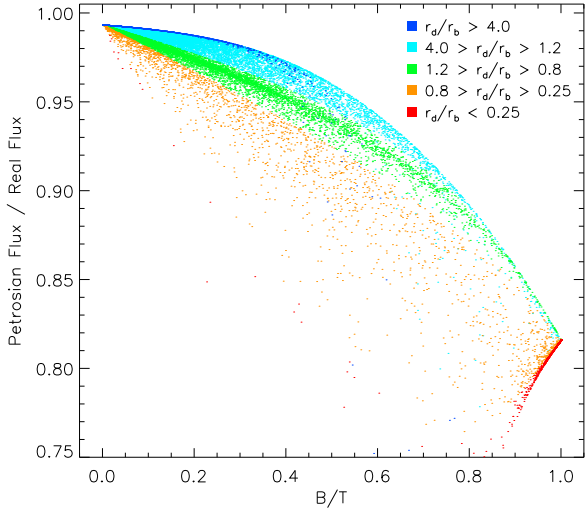}
\caption{
The ratio of Petrosian flux to total flux for a sample of {\tt GALFORM} 
galaxies with the same selection as the SDSS, plotted as a function of the 
bulge to total luminosity ratio (B/T), measured in the $r-$band taking 
into account dust extinction. Points are colour-coded according to the 
ratio of disk to bulge scale-lengths, as indicated by the legend.
}
\label{BTvsPF}
\end{figure}

In this section we describe how standard {\tt GALFORM} outputs, such
as disk and bulge total luminosities and half-mass radii are
transformed into quantities which are measured for SDSS galaxies. This
allows a direct comparison between the model predictions and the
observations. We outline the calculation of Petrosian magnitudes
(\S3.1), the concentration index (Morgan 1958; \S3.2) and the S\'{e}rsic
index \citep{Ser68}. The latter two quantities are used as proxies for
morphological type in analyses of SDSS data. We also illustrate how
the Petrosian magnitude, concentration index and S\'{e}rsic index depend
on the bulge-to-total luminosity ratio and on the ratio of the disk
and bulge radii.

\subsection[]{Petrosian magnitude}

As a measure of galaxy flux, the SDSS team uses a modified definition of 
the Petrosian (1976) magnitude. The Petrosian radius $r_{\rm Pet}$ is 
defined as the radius for which the following condition holds: 

\begin{equation}
\frac{\int_{0.8r_{\rm Pet}}^{1.25r_{\rm Pet}}{\rm d}r2\pi 
rI(r)/[\pi(1.25^2-0.8^2)r_{\rm Pet}^2]}{\int_{0}^{r_{\rm Pet}}
{\rm d}r2\pi rI(r)/[\pi r_{\rm Pet}^2]}=0.2, 
\label{Petr}
\end{equation}
where I(r) is the surface brightness profile. Defined in this way,
$r_{\rm Pet}$ is the radius where the local surface brightness
averaged within a circular annulus centred on the Petrosian radius is
0.2 times the mean surface brightness interior to that radius. The
Petrosian flux defined by the SDSS is then obtained within a circular
aperture of radius $2r_{\rm Pet}$. In the SDSS, the aperture used in
all five bands is set by the profile of the galaxy in the r-band.
I(r) is the azimuthally-averaged surface brightness measured in a
series of annuli. In the case of {\tt GALFORM} model galaxies, we
calculate the disk and bulge sizes, and adopt an exponential profile
for the disk with $I(r) \propto \exp(-1.68(r/r_{\rm D}))$, where
$r_{\rm D}$ is the disk half-light radius, and a de Vaucouleurs
profile for the bulge (assumed to be spherical), with $I(r) \propto
\exp(-7.67(r/r_{\rm B})^{1/4})$, where $r_{\rm B}$ is the bulge
half-light radius in projection (see Cole et~al. 2000). The total
surface brightness profile for a galaxy is given by the sum of the
disk and bulge profiles. The disk and bulge magnitudes include dust
extinction. A random inclination angle is assigned to the galactic
disk for calculating the dust extinction. The Petrosian flux within a
circular aperture of $2r_{\rm Pet}$ recovers a fraction of the total
light of the galaxy which depends on its luminosity profile and hence
its morphology. For a pure disk with an exponential profile, the
Petrosian flux recovers in excess of 99\% of the total flux. On the
other hand, for a pure bulge with a de Vaucouleurs profile, the
percentage of the total light recovered by the Petrosian magnitude is
closer to 80\%. Fig.~\ref{BTvsPF} shows the ratio of Petrosian flux to
total flux for model galaxies as a function the bulge-to-total
luminosity ratio in the $r$-band. The limiting cases described above
are apparent in the plot, which also shows the fraction of light
recovered by the Petrosian definition for composite disk plus bulge
systems, and for different ratios of the disk and bulge scale-lengths.

\subsection{Concentration index}
\begin{figure}
\includegraphics[width=8.5cm]{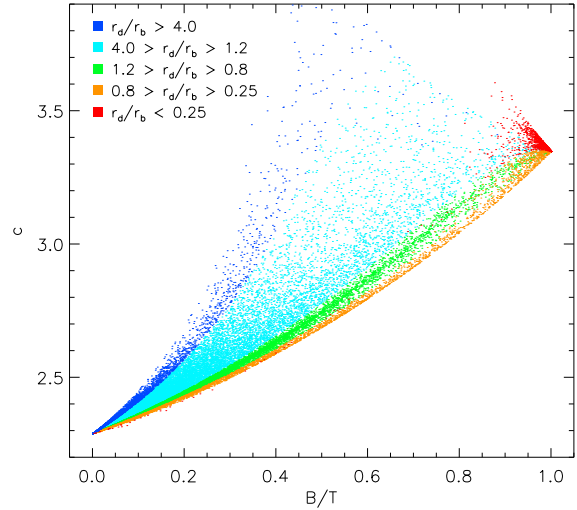}
\caption{
The concentration index, $c$, plotted as a function of the bulge-to-total 
luminosity (B/T) in the $r-$band for the Baugh2005 model galaxies. The 
ratio of disk to bulge scale-lengths is indicated by the colour of the 
symbol as shown by the key. 
}
\label{BTvsc}
\end{figure}

The concentration index can be straightforwardly derived once the
Petrosian flux and radius have been calculated. The concentration
index is defined as $c = R_{90}/R_{50}$, where $R_{90}$ and $R_{50}$
correspond to the radii enclosing $90\%$ and $50\%$ of the Petrosian
flux respectively in the $r$-band. Hence, the luminosity is dominated
by the bulge for high concentration galaxies and is dominated by the
disk for low concentration galaxies. In Fig.~\ref{BTvsc} we plot the
bulge-to-total luminosity versus the concentration index for model
galaxies in the r-band.  We can see that pure disks have $c = 2.3$,
pure bulges have $c = 3.3$, and intermediate values of concentration
index correspond to galaxies with different combinations of $r_{\rm
disk}/r_{\rm bulge}$ and B/T. Most galaxies lie in a narrow locus of
B/T versus c, but different combinations of disk and bulge
scale-lengths and B/T produce the scatter shown.  Observationally, the
Petrosian concentration index is affected by seeing \citep{Bla03}. The
same galaxy can show different concentrations under different seeing
conditions.

\subsection{S\'{e}rsic index}

\begin{figure}
\includegraphics[width=8.7cm]{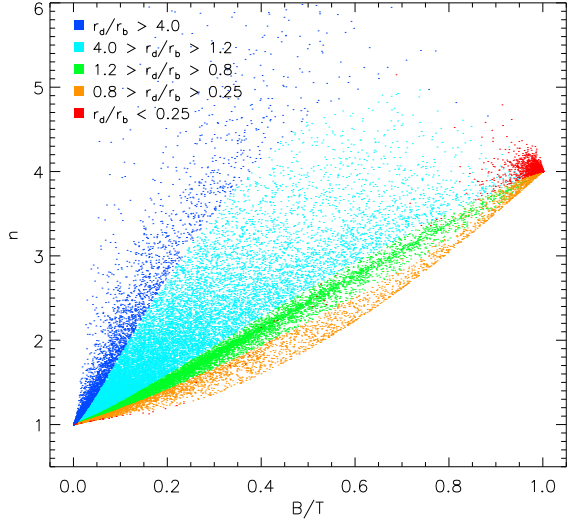}
\caption{
The S\'{e}rsic index $n$ plotted against the bulge-to-total 
luminosity ratio (B/T) for the Baugh2005 model galaxies. 
The ratio of disk to bulge scale-lengths is indicated by 
the colour of the symbol, as shown by the key.
}
\label{B05BTvsN}
\end{figure}

The S\'{e}rsic index describes the shape of a fit made to the 
surface brightness profile of a galaxy without prior knowledge 
of the scale-lengths of the disk and bulge components. The radial 
dependence of the profile is given by (S\'{e}rsic 1968): 

\begin{equation}
I(r)=I_{0}\exp[-(r/r_{0})^{1/n}].
\label{Sers}
\end{equation}
Here $I_{0}$ is the central surface brightness, $r_{0}$ is the
S\'{e}rsic scale radius and $n$ is the S\'{e}rsic index. The S\'{e}rsic
index has been shown to correlate with morphological type
(e.g. Trujillo, Graham \& Caon 2001). We can see that if $n=1$ we
recover an exponential profile (used for pure disk galaxies) and if
$n=4$ we recover a de Vaucouleurs profile, used to describe pure bulge
galaxies. The S\'{e}rsic index has been calculated in the New York
University Value Added Catalogue (NYU-VAGC; \citealt{Bl05}).  Here, we
do not attempt to reproduce exactly the procedure that Blanton
et~al. used to obtain the parameters in Eq.~\ref{Sers} (which takes
into account seeing and pixelization).  Since we know the full surface
brightness profile of the model galaxies out to any radius, we want
the S\'{e}rsic profile that best reproduces the composite disk plus
bulge profile. In order to determine the parameters of the S\'{e}rsic
profile, $I_{0}$, $r_{0}$ and $n$, we minimize a $\chi^2$ function
which depends on the difference between the S\'{e}rsic profile and the
sum of the disk and bulge surface brightness profiles, given the
scale-lengths and luminosity ratio of these components:

\begin{equation}
\chi^{2}=\sum_{i}[\log I_{{\rm disk}+{\rm bulge}}(r_{i})-\log I_{{\rm S\acute{e}rsic}}(r_{i},r_{0},n)]^{2}W_{i}.
\label{Chi2}
\end{equation}
The total luminosity of the fitted S\'{e}rsic profile is constrained to
be equal to that in the true disk + bulge profile. Here $r_{i}$ is a
series of rings between $r=0$ to $r=R_{90,D+B}$, the radius enclosing
the 90\% of the disk plus bulge profile flux, and the weight $W_{i}$
is given by the luminosity in the ring containing $r_{i}$. Since the
steepness of the S\'{e}rsic profile is more evident at the centre of the
galaxy, we assign half of the bins to the region within the bulge size
$r_{\rm bulge}$ (so as long as $r_{\rm bulge} < R_{90,D+B}$). As a
test to check the consistency of changing from a disk+bulge profile to
the S\'{e}rsic profile, we have compared $R_{50}$ (the radius enclosing
50\% of the total luminosity) obtained from the two descriptions of
the surface density profile and find very similar results.

If it is assumed that the distribution of light in real galaxies is
accurately described by the S\'{e}rsic profile, then quantities derived
from fitting a S\'{e}rsic profile have some advantages over the
corresponding Petrosian quantities. (i) The total flux integrated over
the fitted S\'{e}rsic profile would equal the true total flux, unlike
the Petrosian flux, which underestimates the true total value,
especially for bulge-dominated galaxies, as shown in
Fig.~\ref{BTvsPF}. (ii) The effects of seeing can be included in the
S\'{e}rsic profile fitting, so that quantities obtained from the fit
(total flux, scale size $r_0$ and S\'{e}rsic index $n$) are in principle
corrected for seeing effects, unlike the corresponding Petrosian
quantities. However, since the Petrosian quantities are the standard
ones used by the SDSS community, in the rest of this paper we work
with Petrosian magnitude and radius, unless otherwise stated.

Fig.~\ref{B05BTvsN} shows the correlation between S\'{e}rsic index and
the bulge-to-total luminosity ratio in the $r$-band. There is a
considerable scatter between these two proxies or indicators of
morphology, driven by the ratio of the scale-lengths of the disk and
bulge components. For example, galaxies with a S\'{e}rsic index of
$n=4$, usually interpreted as a pure bulge light profile, can have
essentially any value of bulge-to-luminosity ratio from
$B/T=0.1$--$1$. A key feature of this plot is the distribution of disk
to bulge size ratios generated by {\tt GALFORM}. It is possible to
populate other parts of the $n$ - $B/T$ plane in the cases of
extremely large or small values of the ratio of disk to bulge
radii. Without the guidance of a physical model, if a grid of $r_{\rm
d}/r_{\rm b}$ was used instead, the distribution of points would be
even broader than shown in Fig.~\ref{B05BTvsN}. Note that only model
galaxies brighter than $M_{r}-5 \log h = -16$ are included in this
plot.  A similar scatter is seen between these two quantities for real
galaxies, as shown by Fig.~\ref{ALLBTvsN} in the Appendix.

\section[]{RESULTS}

The primary observational data set we compare the model predictions
against is the New York University Value Added Catalogue (NYU-VAGC),
which gives additional properties to those found in the SDSS database
for a subset of DR4 galaxies \citep{Bl05}. The NYU catalogue covers an
area of 4783 deg$^{2}$ and contains 49968 galaxies with redshifts.
The sample is complete to $r_{\rm Pet} = 17.77$ over the redshift
interval $0.0033 < z < 0.05$, and has a median redshift of $z=0.036$.
The relatively low median redshift compared with the full
spectroscopic sample is designed to provide a sample of large galaxy
images, suitable for measurements of galaxy morphology.  Examples of
the extra properties listed for galaxies, beyond the information
available in the SDSS database, include the rest-frame (AB) absolute
magnitude, the S\'{e}rsic index and the value of $V_{\rm max}$ (i.e. the
maximum volume within which a galaxy could have been observed whilst
satisfying the sample selection; this quantity is used to weight each
galaxy in statistical analyses).  We run {\tt GALFORM} with an output
redshift of $z=0.036$ to match the median of the NYU-VAGC and derive
properties from the output which can be compared directly against the
observations, as described in \S 3.  In \S 4.1, we compare the model
and observed luminosity functions, in \S 4.2 we show the distribution
of morphological types versus luminosity, in \S 4.3 we examine the
colour distributions and explore this further as a function of
morphology in \S 4.4. Finally in \S 4.5 we test the size predictions
against observations and assess the sensitivity of the model output to   
the strength of various processes.

\subsection{Luminosity function: all galaxies and by colour}

\begin{figure}
\includegraphics[width=8.5cm]{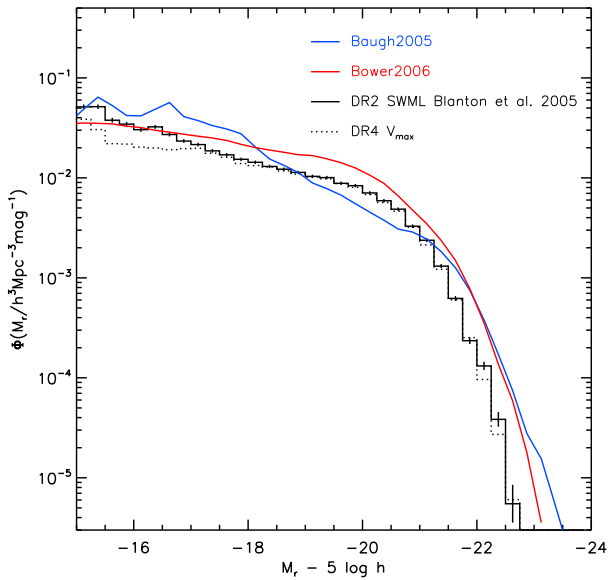}
\caption{
The $r-$band luminosity function predicted by the Bower2006 (red) and 
Baugh2005 (blue) models. For comparison, we also plot the SDSS luminosity 
function estimated using the SWML estimator by Blanton et al (2005) from 
DR2 (solid histogram) and our result using the $1/V_{\rm max}$ estimator 
from DR4 (dotted histogram).
}
\label{LumFun}
\end{figure}

\begin{figure*}
\includegraphics[width=16cm]{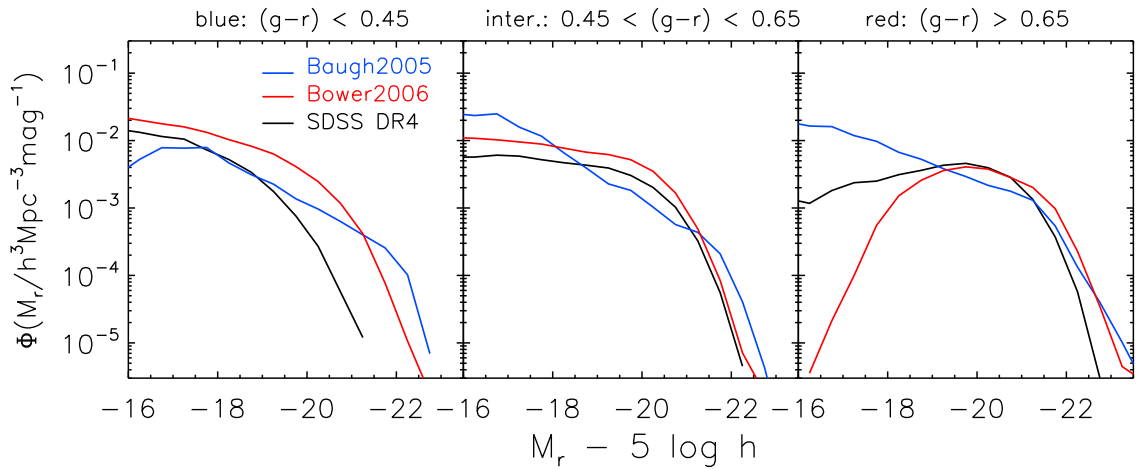}
\caption{
The colour-dependent luminosity function, for three populations of 
galaxies defined by the $g-r$ colour as shown by the label on top 
of each panel. The black lines show the luminosity function estimated 
from DR4 and the red (Bower2006) and blue (Baugh2005) lines show the 
model predictions (see the legend in the left panel).
}
\label{LumFunbyCol}
\end{figure*}

The local luminosity function plays a key role in constraining the
parameters which specify a galaxy formation model. The comparison
between the predicted and observed luminosity functions is hence a
fundamental test of any model. The original papers describing the
Baugh2005 and Bower2006 models showed the comparison of the model
predictions with the observed local luminosity function in the optical
and near-infrared. However a comparison with SDSS data was not made in
those papers. Fig.~\ref{LumFun} shows the luminosity function in the
Petrosian $r$-band predicted by the Baugh2005 and Bower2006 models,
compared with our estimate of the luminosity function from SDSS DR4
made using the values of $V_{\rm max}$ from the NYU-VAGC catalogue. We
also overplot the luminosity function estimated from the SDSS DR2 by
\citet{Bl05b} using the stepwise maximum likelihood (SWML) method. The
SWML and $1/V_{\rm max}$ estimates are in very good agreement,
particularly for magnitudes brighter than $M_{r}-5\log h=-17$. At
fainter magnitudes, the $1/V_{\rm max}$ estimator could be affected by
very local large scale structure \citep{Bl05b}.

Both models overpredict the abundance of bright galaxies.  The
Bower2006 model produces a somewhat better match to the shape of the
SDSS luminosity function.  This offset in the $r-$band luminosity
function has also been noted by Cai et~al. (2008), who made the model
galaxies in the Bower2006 model fainter by 0.15 magnitudes before
using this model to make mock galaxy surveys.  It is worth noting that
the Bower2006 model gives an excellent match to both the $b_{\rm
J}$-band luminosity function estimated from the 2dF galaxy redshift
survey (\citealt{Nor02}) and to the K-band luminosity function
(e.g. Cole et~al. 2001; Kochanek et~al. 2001) without the need to
shift the model magnitudes by hand.

We can study the impact on the luminosity function of different
physical processes in more detail by using colour to separate galaxies
into different samples. For the SDSS data we can compute the
luminosity function of colour sub-samples using the $1/V_{\rm max}$
estimator, bearing in mind that fainter than $M_{r}-5\log h=-17$ this
method gives an unreliable estimate of the luminosity function due to
local large-scale structure. We use the Petrosian $g-r$ colour to
split galaxies into blue ($g-r < 0.45$), red ($g-r > 0.65$) and
intermediate ($0.45 < g-r < 0.65$) colour samples. In
Fig.~\ref{LumFunbyCol}, we can see that both models reproduce the
intermediate colour population fairly well (middle panel). The
Bower2006 model in particular matches the shape of the observed
luminosity function closely, albeit with a shift to brighter
magnitudes, similar to that seen in the case of the overall luminosity
function in Fig.~\ref{LumFun}. The models fare worst for blue
galaxies, with both models overpredicting the number of bright blue
galaxies.  This suggests that star formation is not quenched
effectively enough in massive haloes or that the timescale for gas
consumption in star formation is too long. For the case of red
galaxies, the models do best brightwards of $L_{*}$, but get the
number of faint red galaxies wrong, with the Baugh2005 model giving
too many faint red galaxies and Bower2006 too few.

\subsection{The distribution of morphological types}

\begin{figure*}
\includegraphics[width=16cm]{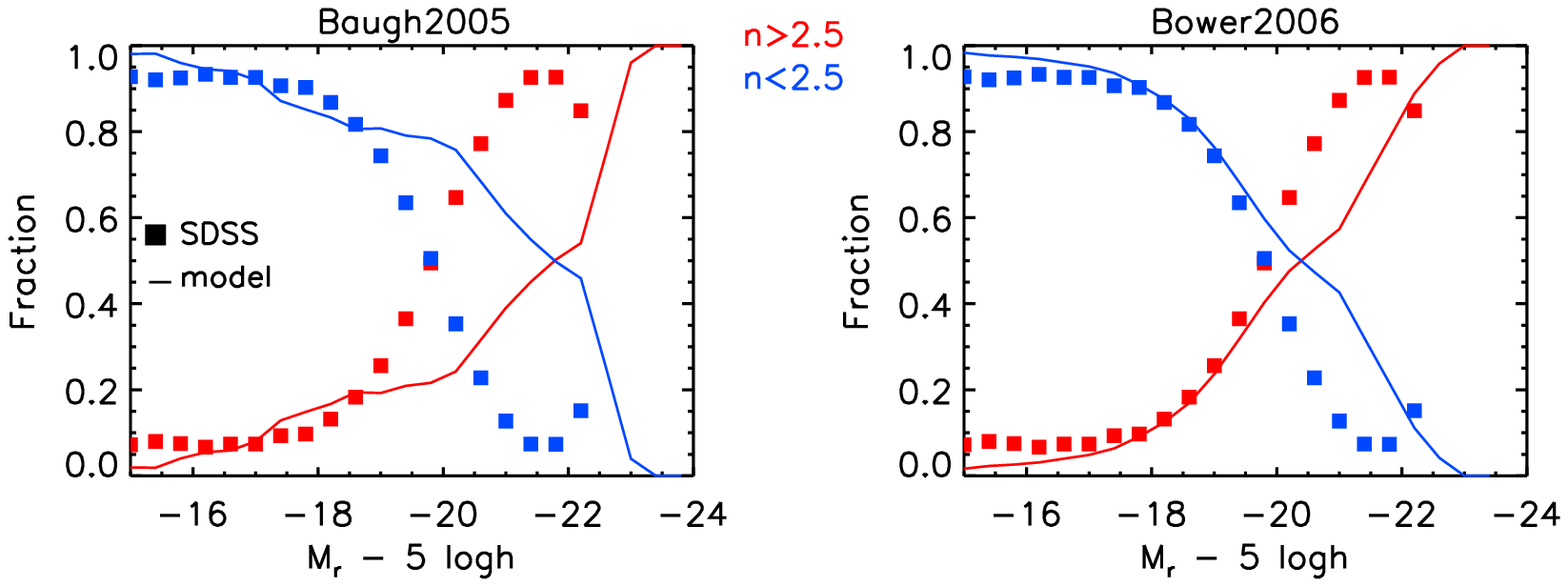}
\caption{
The fraction of different morphological types as a function of magnitude 
$M_{r}$ for SDSS data (squares) and {\tt GALFORM} (solid lines). 
The fraction of disk-dominated galaxies (as defined by a value of the 
S\'{e}rsic index $n<2.5$) is shown in blue and bulge-dominated 
galaxies (i.e. those with $n>2.5$) are plotted in red. The left panel shows 
the Baugh2005 model and the right panel the Bower2006 model.
}
\label{Morfn}

\includegraphics[width=16cm]{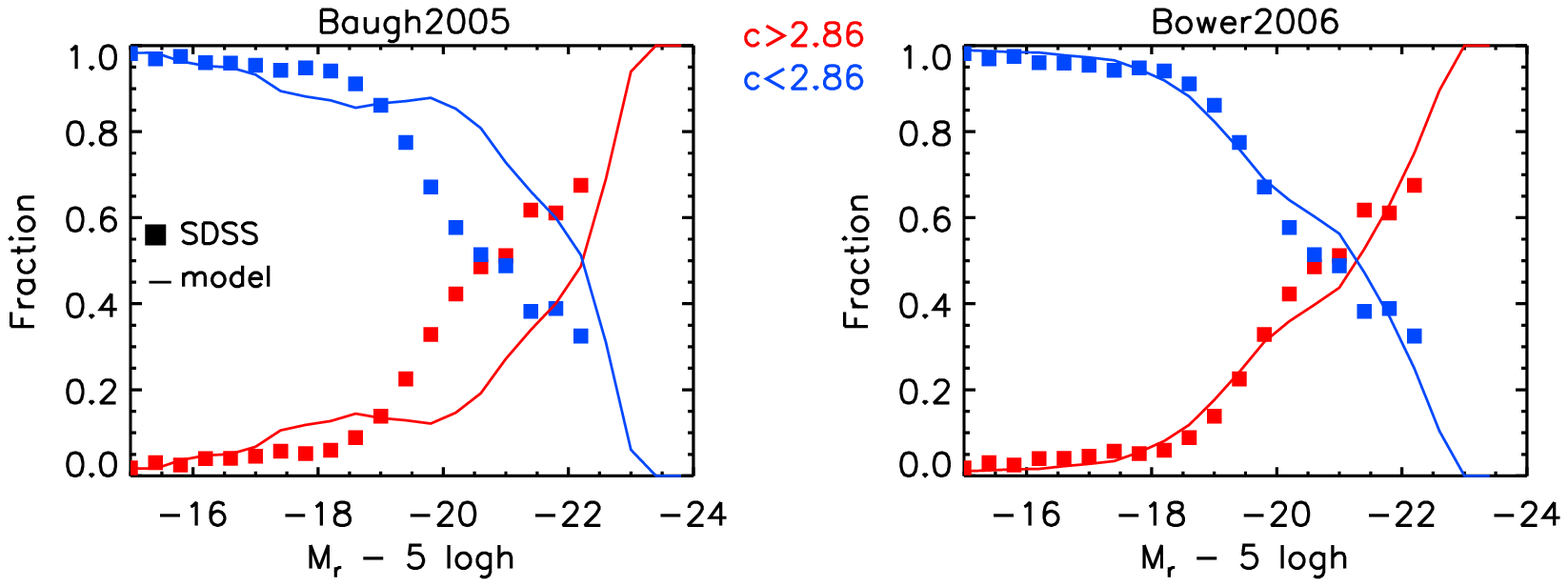}
\caption{ The fraction of different morphological types as a function
of magnitude $M_{r}$ for SDSS data (squares) and {\tt GALFORM} (solid
lines), using the Petrosian concentration index $c$ to define type.
The fraction of disk-dominated galaxies (as defined by $c<2.86$) is
shown in blue and bulge-dominated galaxies (i.e. those with $c>2.86$)
are plotted in red. The left panel shows the Baugh2005 model and the
right panel the Bower2006 model.  }
\label{Morfc}

\includegraphics[width=16cm]{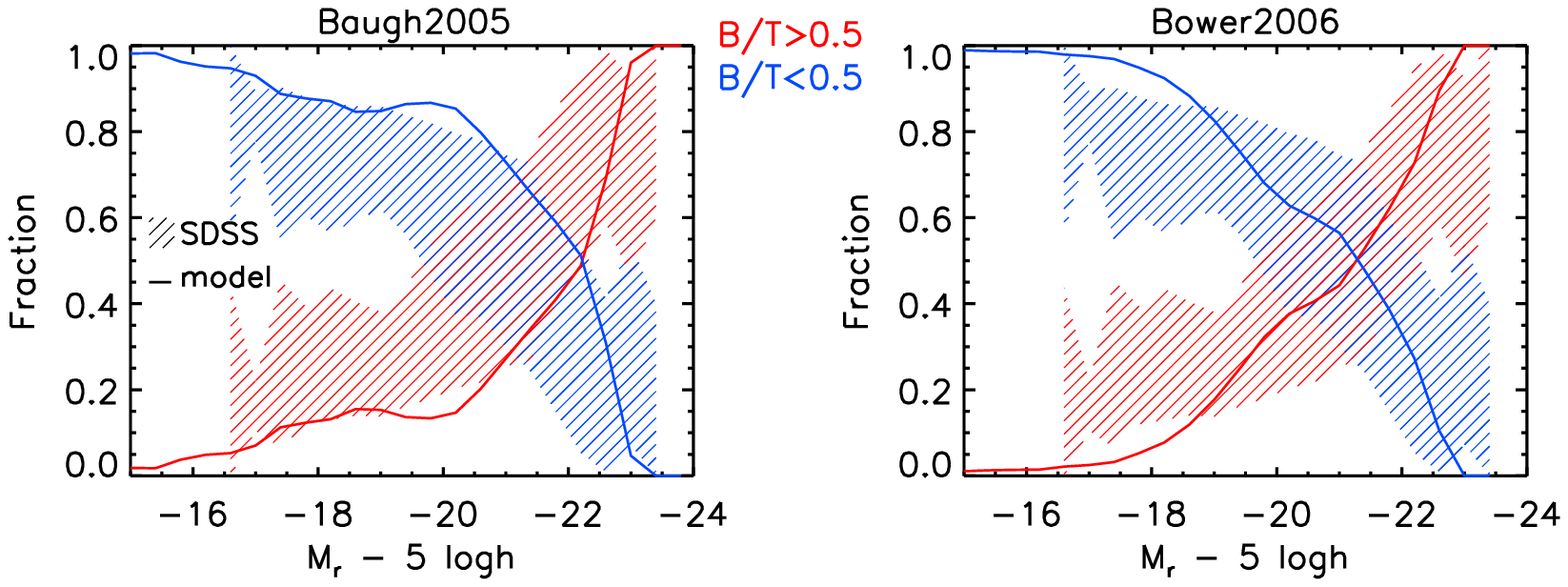}
\caption{
The fraction of different morphological types as a function of magnitude 
$M_{r}$ using the bulge-to-total luminosity 
ratio in the $r-$band to define type. Disk-dominated galaxies ($B/T<0.5$) 
are shown in blue and bulge-dominated ($B/T>0.5$) are plotted in red. 
The solid curves show the model predictions, according to the label above 
each panel. The shaded region shows an observational estimate made from 
SDSS data by Benson et~al. (2007). The extent of the shading shows by how much 
the fraction changes when a correction is applied to the observational 
estimates (see Benson et~al. for details). 
}
\label{MorfBT}
\end{figure*}

In this section we examine the mix of morphological types as a function 
of luminosity, using different proxies for galaxy morphology. 

We first look at the mix of galaxies using the S\'{e}rsic index.  Using
a S\'{e}rsic index value of $n=2.5$ (which is half-way between $n=1$
and $n=4$), we separate galaxies into two broad morphological classes,
disk-dominated galaxies (late type galaxies) with $n<2.5$ and bulge-dominated
galaxies (early type galaxies) with $n>2.5$.
Fig.~\ref{Morfn} shows the fraction of galaxies in each morphological
type, as a function of Petrosian magnitude, $M_{r}$, for SDSS galaxies
and the {\tt GALFORM} model (Baugh2005 in the left panel and Bower2006
in the right panel). The trend found for SDSS galaxies is that the
disk-dominated population is the more common at faint magnitudes,
whereas bulge-dominated objects are in the majority brighter than
$L_{*}$. Fig.~\ref{Morfn} shows that both models follow the same
general trend, but with the changeover from one population to the
other occuring brighter than $L_{*}$ in the Baugh2005 model, whereas
the Bower2006 model looks more similar to the observations.

Another way to morphologically classify galaxies using the profile
shape is to use the Petrosian concentration index $c$. In
Fig.~\ref{Morfc}, we show the fraction of early and late type galaxies
as a function of luminosity based on this, where we classify galaxies
with $c<2.86$ as late type and $c>2.86$ as early type. The resulting
plots look very similar to those based on S\'{e}rsic index in
Fig.~\ref{Morfn}, though there are differences in detail, particularly
at the bright end. The agreement between the models and the SDSS data
is generally better using the concentration as the classifier,
particularly for the Bower2006 model.

As a third approach to determining galaxy type, we consider the
bulge to total luminosity ratio, $B/T$ measured in the $r-$band.
\citet{Ben07} fitted disk and bulge components to images of 8839
bright galaxies selected from the SDSS EDR. In fitting the disk and
bulge components of each galaxy, they used the bulge ellipticity and
disk inclination angle, $i$, as free parameters. The resulting
distribution of $\cos(i)$ showed an excess of face-on galaxies. This
is due in part to the algorithm mistaking part of the bulge as a
disk. Benson et~al. attempted to correct for the uneven distribution
of inclination angles in the following way. Galaxies with
$\cos(i)<0.5$ are assumed to have been correctly fitted. Since a
uniform distribution in $\cos(i)$ is expected, for each galaxy with
$\cos(i)<0.5$, a galaxy with a similar bulge and face-on, projected
disk magnitude but with $\cos(i)>0.5$ is also selected.  The galaxies
with $\cos(i)>0.5$ which are left without a match are assumed to
correspond to cases where the disk component has been used to fit some
feature in the bulge. Benson et~al. assigned to these galaxies a value
of B/T=1. The correction has a considerable impact on the fraction of
bulge- and disk-dominated galaxies, as shown by the extent of the
shaded region in Fig.~\ref{MorfBT}.

The observational estimates of the mix of morphological types
presented in Figs.~\ref{Morfn}, \ref{Morfc} and \ref{MorfBT} are
qualitatively the same, but show that the transition from
disk-dominated to bulge-dominated depends on the choice of property
used to define morphology. We note that the model predictions 
are very similar when we set the division at $c=2.86$ or at $B/T=0.5$.
The model predictions made using the
S\'{e}rsic index, concentration and bulge-to-total luminosity ratio
appear to be closer to each other than the corresponding observational measurements.
This comparison gives some indication of the observational uncertainty
in measuring fractions of different morphological types using the
S\'{e}rsic index, concentration and bulge-to-total luminosity ratio.

\subsection{Colour distribution}

\begin{figure*}
\includegraphics[width=14.7cm]{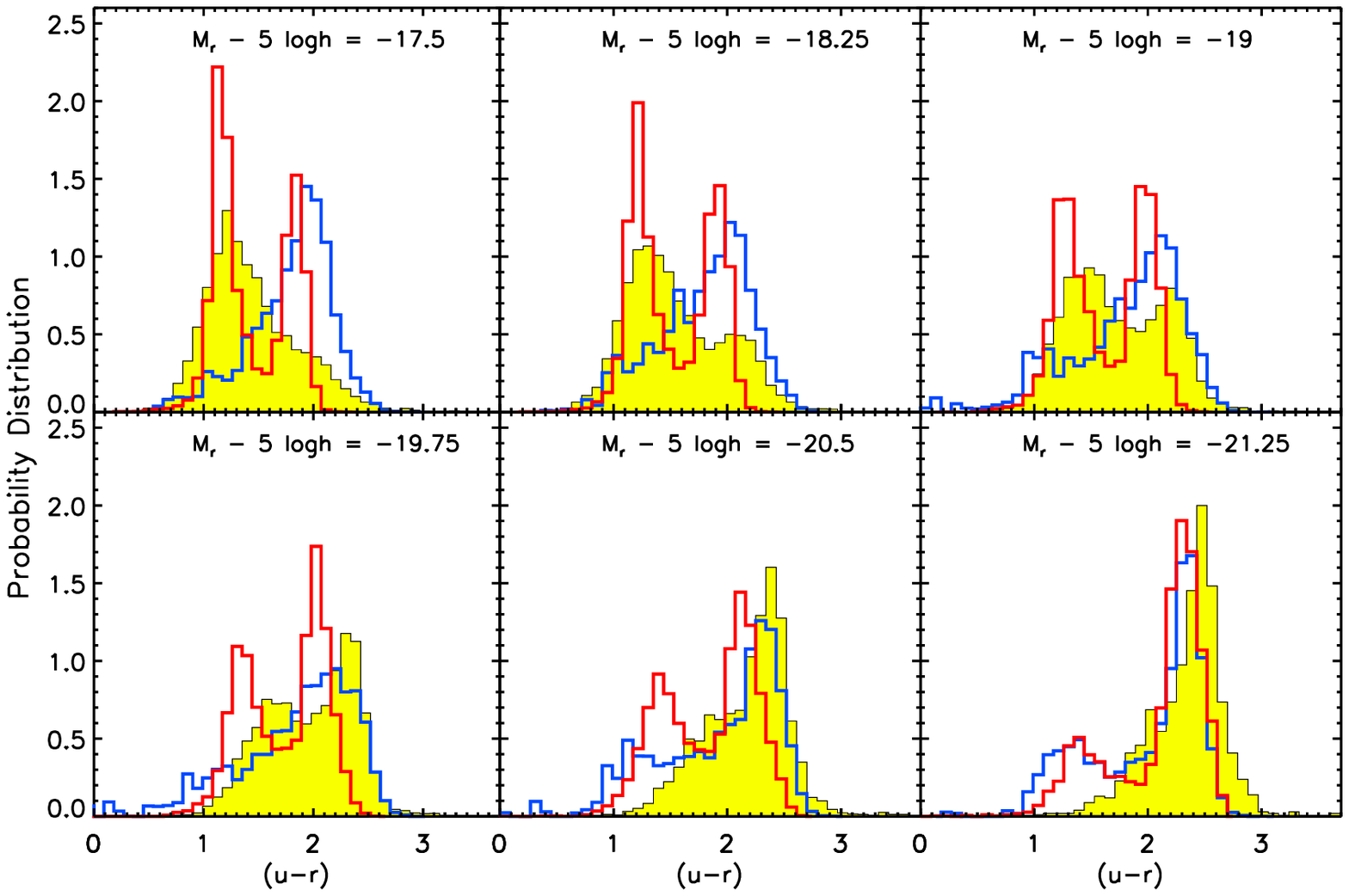}
\caption{
The $(u-r)$ colour distribution as function of luminosity for the 
Baugh2005 model (blue histograms), the Bower2006 model (red histograms) 
and the SDSS data (yellow histograms). The centre of the magnitude bin 
used in each panel is given by the legend.
All the histograms are normalized to have unit area. 
}
\label{Colorshort}
\end{figure*}

\begin{figure*}
\includegraphics[width=12.5cm]{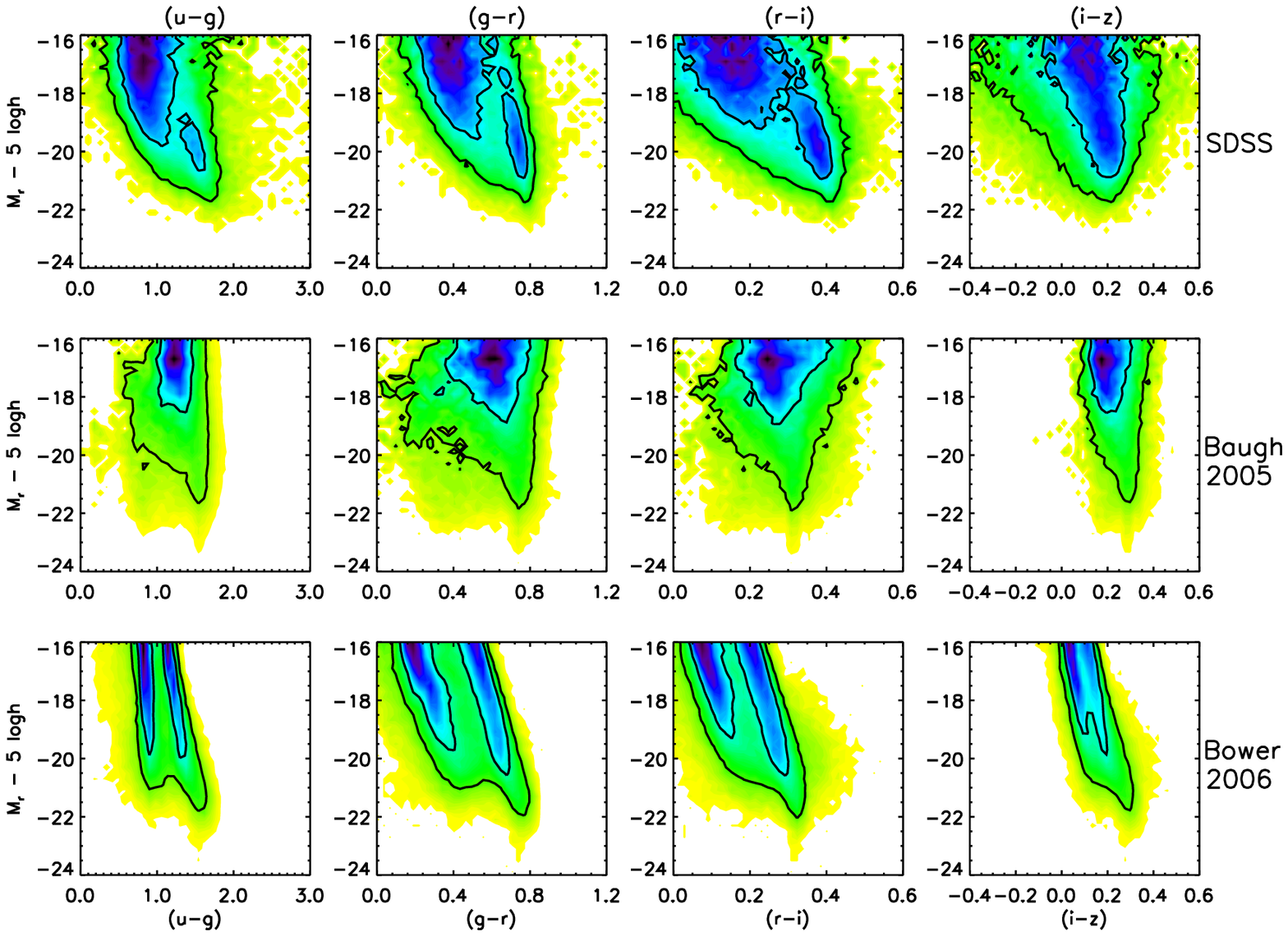}
\caption{
The galaxy distribution in the colour magnitude plane. Each column shows the 
distribution for a different colour. The top row shows the SDSS distributions, 
the middle row the predictions of the Baugh2005 model and the bottom row the 
Bower2006 model. Galaxies are weighted by $1/V_{\rm max}$. The inner contour 
encloses 68\% of the total number density of galaxies and the outer contour 
encloses 95\% of the density. The colour shading reflects the square root 
of the number density. 
}
\label{ColorvsmW}
\includegraphics[width=12.5cm]{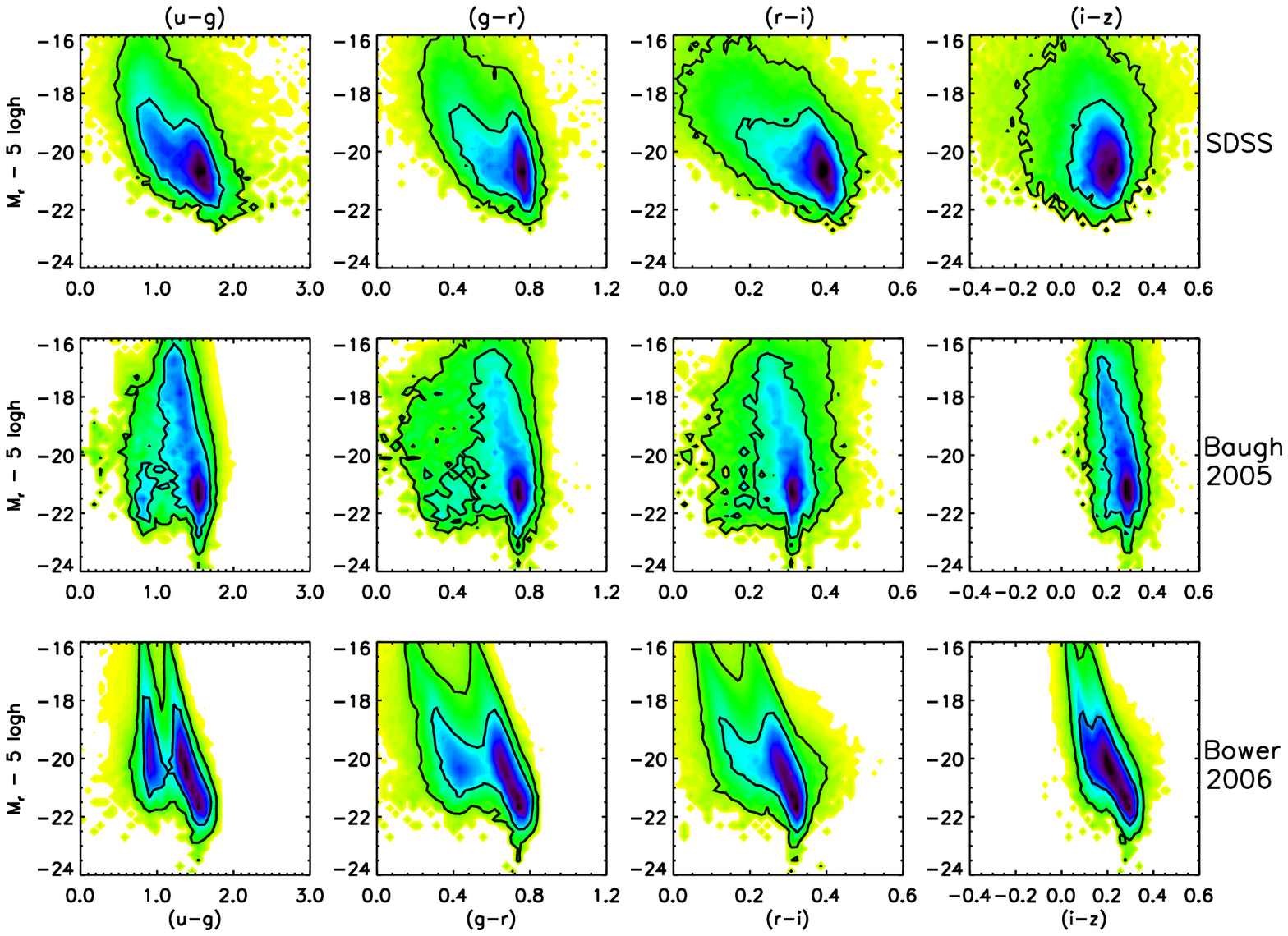}
\caption{
The same as Fig.~\ref{ColorvsmW}, but with each galaxy weighted by the 
product of its luminosity and $1/V_{\rm max}$. 
}
\label{ColorvsmLW}
\end{figure*}

An important feature uncovered in the SDSS data is a bimodality in the
galaxy colour-magnitude relation (e.g. \citealt{Str01};
\citealt{Bal04}).  In Fig.~\ref{Colorshort} we plot the distribution
of Petrosian $(u-r)$ colour in selected bins of magnitude $M_{r}$ for
models and SDSS data. 
The SDSS data shows a dominant red population at
bright magnitudes, with a blue population that becomes more important
at fainter magnitudes.  Although we see blue and red populations in
the Baugh2005 model predictions for intermediate magnitudes, the red
population always dominates, even at the faintest magnitudes. The
Bower2006 model, on the other hand displays a clear bimodality, with
the red population dominating at bright magnitudes, comparable red and
blue populations at intermediate magnitudes and a slightly more
dominant blue population at faint magnitudes. The Bower2006 model
shows the same behaviour as the SDSS data at bright magnitudes. At
faint magnitudes, the Bower2006 model still shows a red population
which is not apparent in the data. \citet{Font08} argued that these
faint red galaxies are predominantly satellite galaxies, which in the
Bower2006 model have exhausted their cold gas reservoirs. In the
\citeauthor{Font08} model, which is a modified version of the
Bower2006 model, the stripping of hot gas from satellites is
incomplete, and so gas may still cool onto the satellite, fuelling
further star formation and causing these galaxies to have bluer
colours on average.

Since the SDSS photometry covers five bands ($u,g,r,i$ and $z$), we
can investigate the bimodality further using different colours.  In
Fig.~\ref{ColorvsmW} we plot the abundance of galaxies in the colour
magnitude plane, for $(u-g),(g-r),(r-i)$ and $(i-z)$ Petrosian colours
against magnitude $M_{r}$. The top row of panels shows the
distributions for the NYU-VAGC SDSS data, the middle row shows the
predictions of the Baugh2005 model, and the bottom row gives those of
the Bower2006 model. In this plot, each galaxy contributes $1/V_{\rm
max}$ to the density.  The contours in the plot indicate the regions
containing 68\% and 95\% of the number density of galaxies in the
samples. Note that the colour shading scales as the square root of the
density.

For the case of the SDSS data in all colours except for $(i-z)$ we can
see a bright red population and a fainter, bluer population as
indicated by the splitting of the 68\% density contour. The Baugh2005
model predicts a dominant red population at the brightest
magnitudes. On moving to fainter magnitudes, bluer galaxies appear but
red galaxies still dominate and there is no clear bimodality. The
Bower2006 model displays a strong bimodality in colour, though with a
bluewards shift in the locus of the colour-magnitude relation compared
with the observations. \citet{Font08} obtained better agreement of
their model with the observed locus of red galaxies in the SDSS by
increasing the assumed yield of metals by a factor of two relative to
the Bower2006 model.
 
In Fig.~\ref{ColorvsmLW} we plot a similar colour-magnitude
distribution, but this time each galaxy contributes $L/V_{\rm max}$ to
the density. By doing this more emphasis is given to brighter
galaxies. As a consequence, the bimodality in the SDSS data is less
readily apparent. Intriguingly, the Baugh2005 model appears visually
to be in better agreement with the observations when presented in this
way.

\subsection{Colour distribution by morphology}

\begin{figure*}
\includegraphics[width=12.7cm]{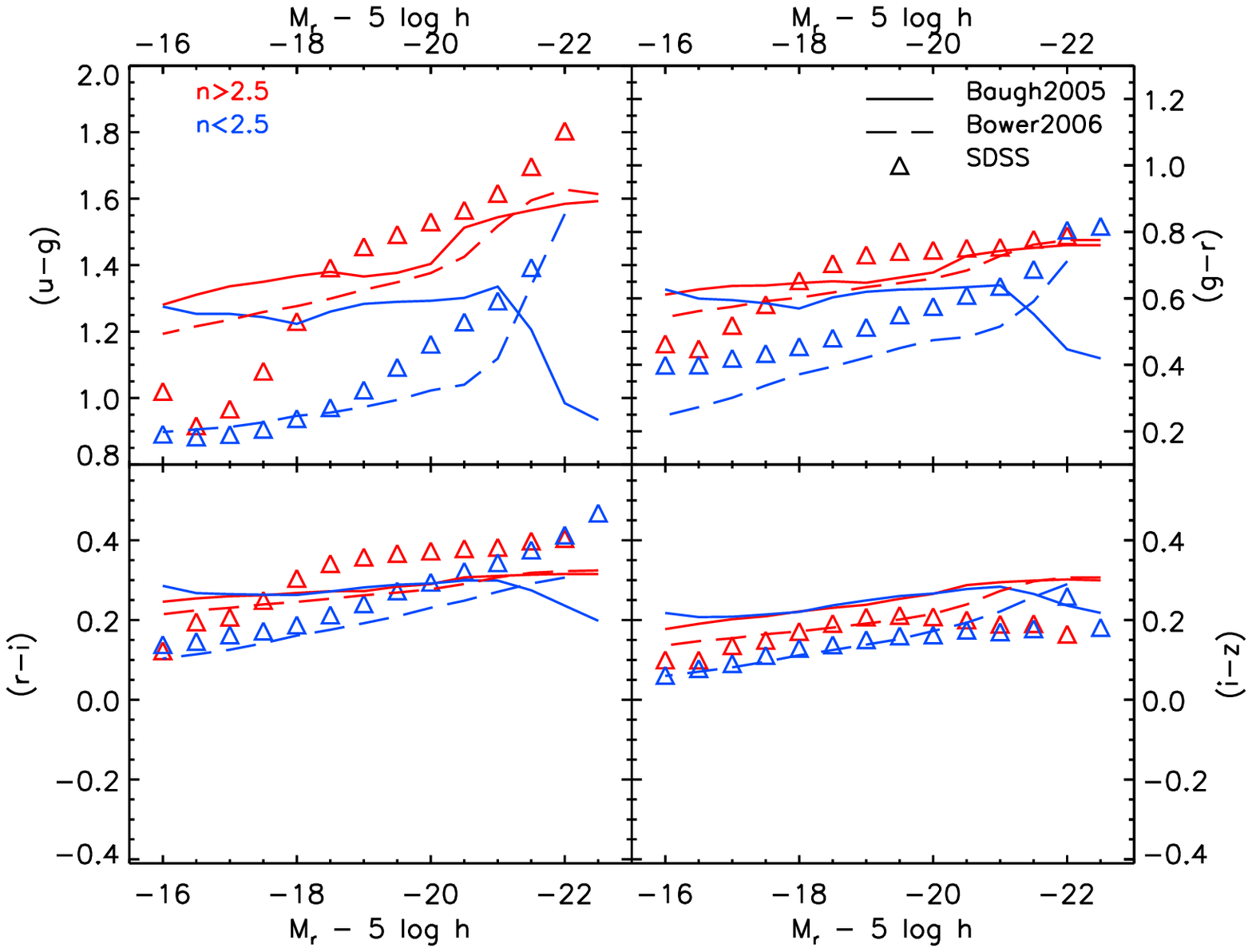}
\caption{
The median colour $(u-g), (g-r), (r-i), (i-z)$ for the models 
(continuous lines) and SDSS galaxies (triangles) as a function 
of magnitude $M_{r}$. Different coloured lines and symbols correspond 
to different morphological types of galaxies, as given by the 
S\'{e}rsic index ($n<2.5$ for the disk-dominated population and $n > 2.5$ 
for the bulge-dominated population). Each panel corresponds to a different 
colour. The panels each have the same range of colour on the y-axis. 
The results are plotted only when there are ten or more galaxies present 
in a bin.
}
\label{Colorvsn}
\end{figure*}

The bimodality of the colour-magnitude relation seen for SDSS galaxies
suggests that different populations or types of galaxy dominate at
different magnitudes. We also saw in Section 4.2 that disk-dominated
galaxies are more abundant at faint magnitudes and the bulge-dominated
population is more prevalent at bright magnitudes. A correlation is
therefore expected between morphology, colour and luminosity. To see
this effect more clearly, we use the S\'{e}rsic index, $n$, to separate
galaxies into an ``early-type'' bulge-dominated population (with
$n>2.5$) and a ``late-type'' disk-dominated population ($n<2.5$) and
replot the colour-magnitude relation.

We calculate the median Petrosian colour for $(u-g),(g-r),(r-i),(i-z)$
in bins of magnitude $M_{r}$ for the two populations. The results are
plotted in Fig.~\ref{Colorvsn} for the models and the SDSS data.  We
can see from the SDSS data that the different populations display
different colour-magnitude correlations, confirming that the S\'{e}rsic
index is an effective morphological classifier. The bulge-dominated
galaxies are redder than the disk-dominated galaxies, with the size of
the difference decreasing as the effective wavelength of the passbands
increases. Also, at fainter magnitudes, the colours of the two
population tend to become more similar.

In the Baugh2005 model, Fig.~\ref{Colorvsn} shows that the populations
split by S\'{e}rsic index have similar colours except for the brightest
galaxies. Both populations are predicted to be too red at faint
magnitudes. At brighter magnitudes ($M_{r}-5\log h<-19$),
bulge-dominated galaxies show similar behaviour to the SDSS
data. Disk-dominated galaxies become bluer at the brightest
magnitudes, which is opposite to the trend seen in the data.  The
Bower2006 model predicts a clear separation in colour for populations
classified by S\'{e}rsic index, with blue disk-dominated galaxies even
at faint magnitudes, which is in better agreement with SDSS data.  In
both models, the faint bulge-dominated population is predicted to be
too red.

\subsubsection{What drives the colours? A look at the specific star 
formation rate and metallicity}

\begin{figure}
\includegraphics[width=8.5cm]{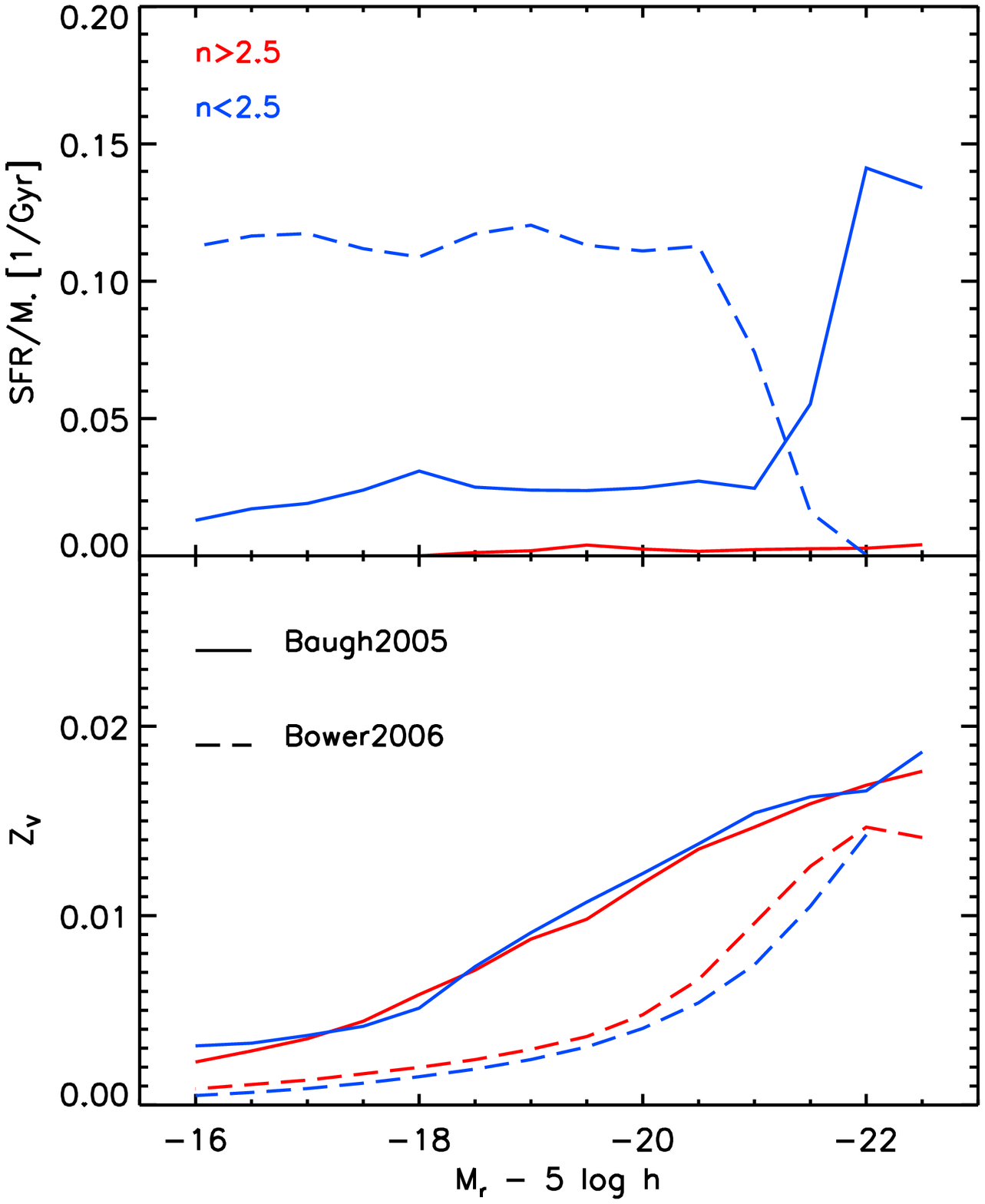}
\caption{
The top panel shows the specific star formation rate, i.e. the star formation 
rate per unit stellar mass, as a function of magnitude $M_{r}$ for the 
{\tt GALFORM} models, as indicated by the key in the lower panel. 
The lower panel shows the stellar metallicity, weighted by the V-band 
luminosity as a function of magnitude $M_{r}$. Different colours correspond 
to different morphological types as set by the S\'{e}rsic index, as shown 
by the label in the top panel. 
}
\label{sfrMetal}
\end{figure}

To identify which feature of the models is producing the differences
in colour seen in Fig.~\ref{Colorvsn}, we now examine the specific
star formation rate SSFR (the star formation rate [SFR] per unit
stellar mass) and stellar metallicity of galaxies in both models. The
specific star formation rate quantifies how vigorously a galaxy is
forming stars in terms of how big a contribution recent star formation
makes to the total stellar mass. Galaxies with a high specific star
formation rate will tend to have bluer colours and stronger emission
lines than more ``passive'' galaxies. We use the S\'{e}rsic index to
separate the galaxies as before, into an ``early-type''
bulge-dominated population (with $n>2.5$) and a ``late-type''
disk-dominated population ($n<2.5$). In the top panel of
Fig.~\ref{sfrMetal} we plot the median of the SSFR as a function of
magnitude $M_{r}$. In the bottom panel of this figure we plot the
median of the V-band luminosity-weighted stellar metallicity.  The top
panel of Fig.~\ref{sfrMetal} shows that bulge-dominated galaxies have
very low specific star formation rates in both models.  The
disk-dominated galaxies have very different specific star formation
rates in the two models. In the Bower2006 model, the disk-dominated
galaxies are undergoing significant amounts of star formation, except
at the brightest magnitudes. Although the strength of supernova
feedback is stronger in the Bower2006 model than it is in the
Baugh2005 model, reheated gas tends to recool on a shorter timescale
because it is reincorporated into the hot halo faster. The drop in the
specific star formation rate for the brightest galaxies in the
Bower2006 model can be traced to the AGN feedback which shuts down gas
cooling for these galaxies. Note that disk-dominated galaxies make up
only a small fraction of the galaxies at these magnitudes.  Within a
given model, the metallicities of the disk and bulge-dominated
populations are similar. However, the metallicities in the Baugh2005
model are higher than in Bower2006, presumably because some fraction
of the star formation in the former model occurs in starbursts with a
top-heavy IMF, which correspondingly produces a higher yield of
metals.  Hence, given this differences, one expects bluer galaxies at
faint magnitudes in the Bower2006 model than in the Baugh2005 model.

\subsubsection{Correlation between S\'{e}rsic index, colour and magnitude}

\begin{figure}
\includegraphics[width=8.5cm]{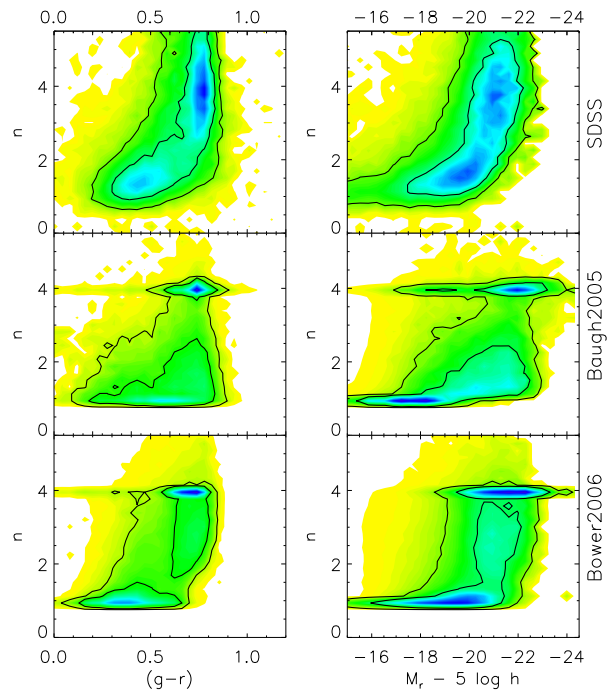}
\caption{
The luminosity-weighted density of galaxies in different 
projections of the S\'{e}rsic index ($n$), (g-r) colour and magnitude plane. 
Contours indicate the regions containing $68\%$ and $95\%$ of the total 
density of galaxies. Top panels: SDSS data, intermediate panels: 
Baugh2005 model and bottom panels: Bower2006 model.
}
\label{nvsM}
\end{figure}

To investigate further the correlation between the S\'{e}rsic index, 
colour and absolute 
magnitude, we plot in Fig.~\ref{nvsM} the luminosity 
weighted density in the various projections of the S\'{e}rsic index ($n$), 
$(g-r)$ colour and $M_{i}$ absolute magnitude plane, both for 
SDSS data and {\tt GALFORM} models. In the data we can see that 
disk-dominated galaxies (i.e. those with small $n$ values) tend to be 
bluer and also fainter, whereas the bulge-dominated galaxies 
(those with large $n$ values) tend to be redder and brighter.
The predictions of both {\tt GALFORM} models are peaked around 
S\'{e}rsic indices of $n=1$ (nominally pure disc galaxies) and $n=4$ 
(pure bulge galaxies). Despite these density peaks, the numbers 
of galaxies in the different morphological classes are similar to 
the SDSS data (as shown in Fig.~\ref{Morfn}) showing that at a broad-brush 
level, the distribution of $n$ predicted by the {\tt GALFORM} models is 
in reasonable agreement with the observations. 

Bearing in mind the level at which the models are able to match the 
distribution of S\'{e}rsic indices, both models reproduce fairly well the 
behaviour seen in the SDSS observation, with a spike corresponding to 
a faint blue disk-dominated population, which changes to a red,  
bulge-dominated population at bright magnitudes. Compared with the SDSS, 
the Baugh2005 model overpredicts the number of red disk-dominated 
galaxies (around values $(g-r) \sim 1$ and $n \sim 1$) and the number 
of moderate luminosity bulge-dominated galaxies (around values 
$M_{i}-5\log h \sim -19$ and $n \sim 4$). The Bower06 model predicts 
a distribution which agrees better with the observational data.

\subsection{The distribution of disk and bulge sizes}

We now examine the model predictions for the linear size of the disk
and bulge components of galaxies. We compare the model predictions
with SDSS observations using the radius enclosing $50\%$ of the
Petrosian flux, $R_{50}$. The calculation of disk and bulge sizes was
reviewed in Section 2.3 (see also Cole et al. 2000 and Almeida et
al. 2007). We use the concentration index, $c$, to divide galaxies
into two broad classes of disk-dominated and bulge-dominated samples.
First we discuss the accuracy of the predictions for $R_{50}$ for
disk-dominated galaxies (\S~4.1) and then for bulge-dominated galaxies
(\S~4.2), before illustrating the sensitivity of the results to
various physical ingredients of the models. The observations we
compare against are our own analysis of the size distribution in the
NYU-VAC constructed from DR4, as discussed below, and the results from
Shen et~al.  (2003; hereafter Sh03), which were derived from a sample
of $~140 000$ galaxies from DR1.

\subsubsection{Disk-dominated galaxies}

\begin{figure*}
\includegraphics[width=18cm]{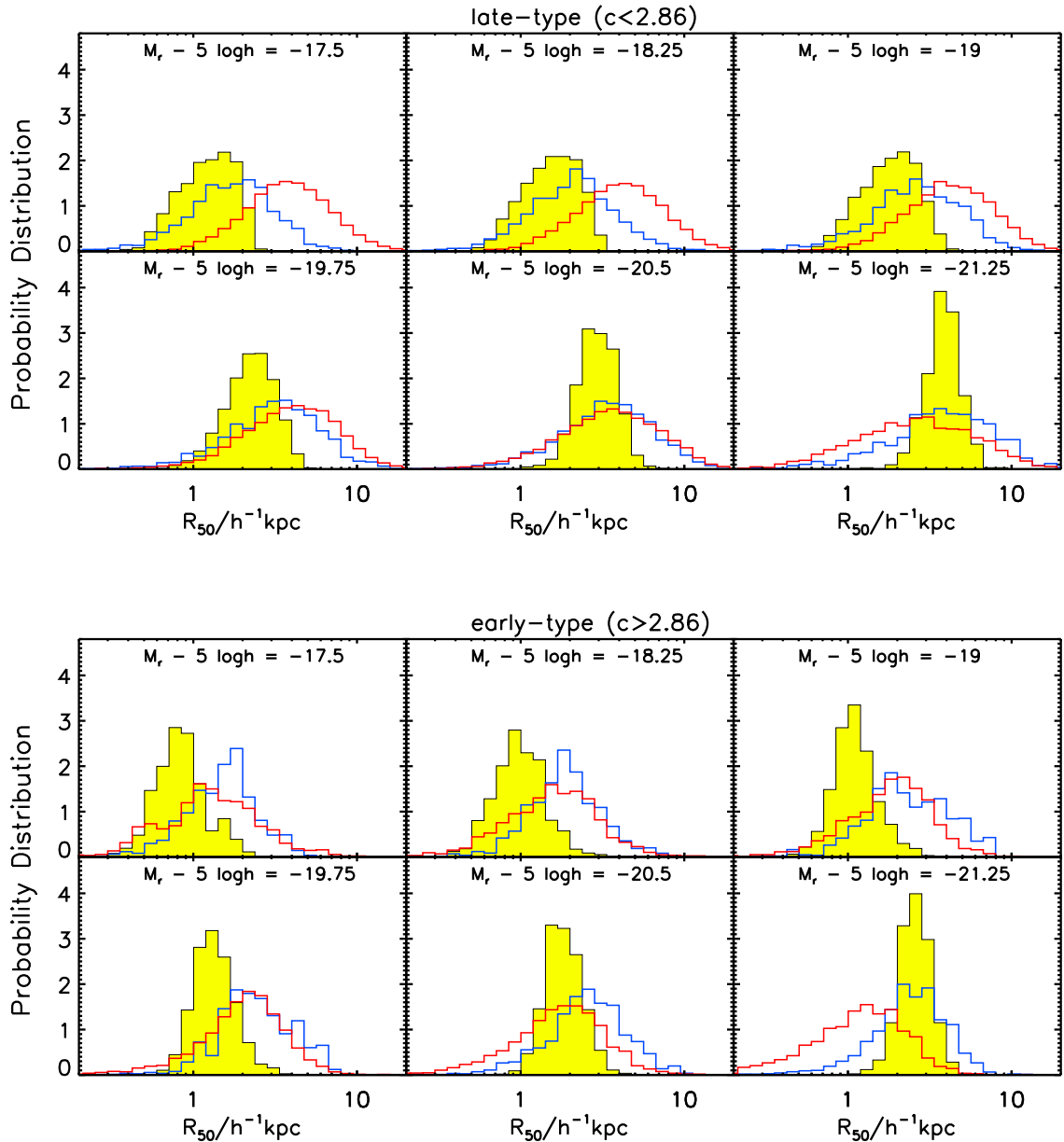}
\caption{ The distribution of the Petrosian half-light radius,
$R_{50}$, for early type galaxies (with concentration parameter
$c>2.86$) in top six panels and for late type galaxies ($c<2.86$) in
the lower six panels. Each panel corresponds to a different one
magnitude wide bin, as indicated by the legend.  The {\tt GALFORM}
predictions are shown by unshaded histograms (Baugh2005 - blue;
Bower2006 - red) and the SDSS data by the yellow shaded histogram. All
of the histograms are normalized to have unit area.}
\label{AllSizesShort}
\end{figure*}

\begin{figure*}
\includegraphics[width=17cm]{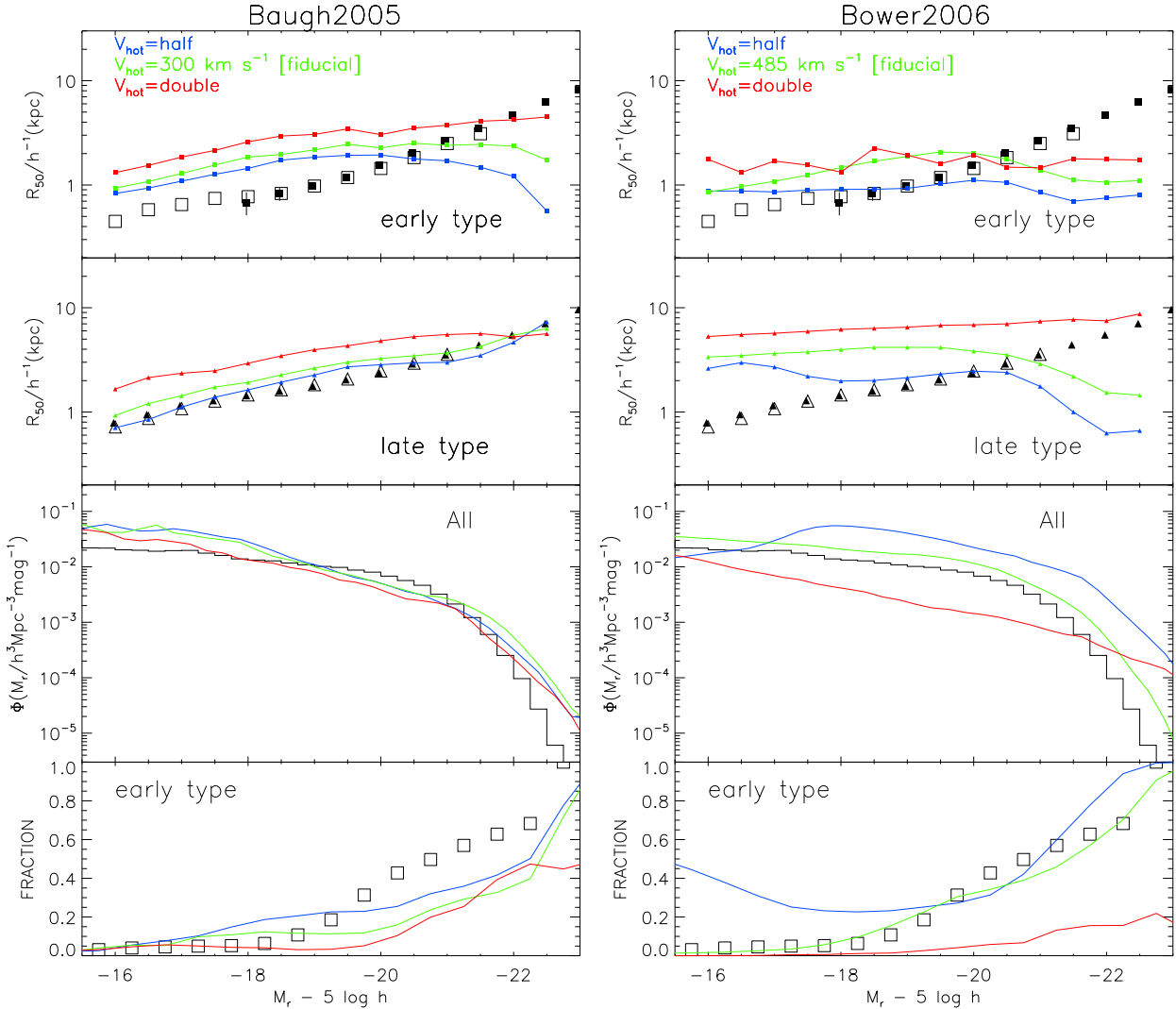}
\caption{ A compilation of predictions for the Baugh2005 (left column)
and Bower2006 (right column) models. The top row shows the median
$R_{50}$ as a function of magnitude for bulge-dominated galaxies
($c>2.86$), the second row shows the median $R_{50}$ for
disk-dominated galaxies ($c < 2.86$), the third row shows the $r-$band
luminosity function of all galaxies and the bottom row shows the
fraction of early-type galaxies as a function of magnitude. The
predictions of the fiducial model in both cases are shown by the green
lines. In this plot, we also show the impact of changing the strength
of supernova feedback, rerunning the model with either half the
fiducial value of $V_{\rm hot}$ (blue curves; see Section 2.1) or
twice the value (red curves).  In first and second rows, the open
symbols show our determination of the median size from the NYU-VAGC;
the filled symbols show the results obtained by Sh03. The black
histogram in the third row shows our determination of the luminosity
function in DR4 using the $1/V_{\rm max}$ estimator. The squares in
the bottom row show the fraction of early-types in the NYU-VAGC,
defined according to concentration parameter $c > 2.86$, as a function
of magnitude.  }
\label{Shenmedvhot}
\end{figure*}

\begin{figure*}
\includegraphics[width=17cm]{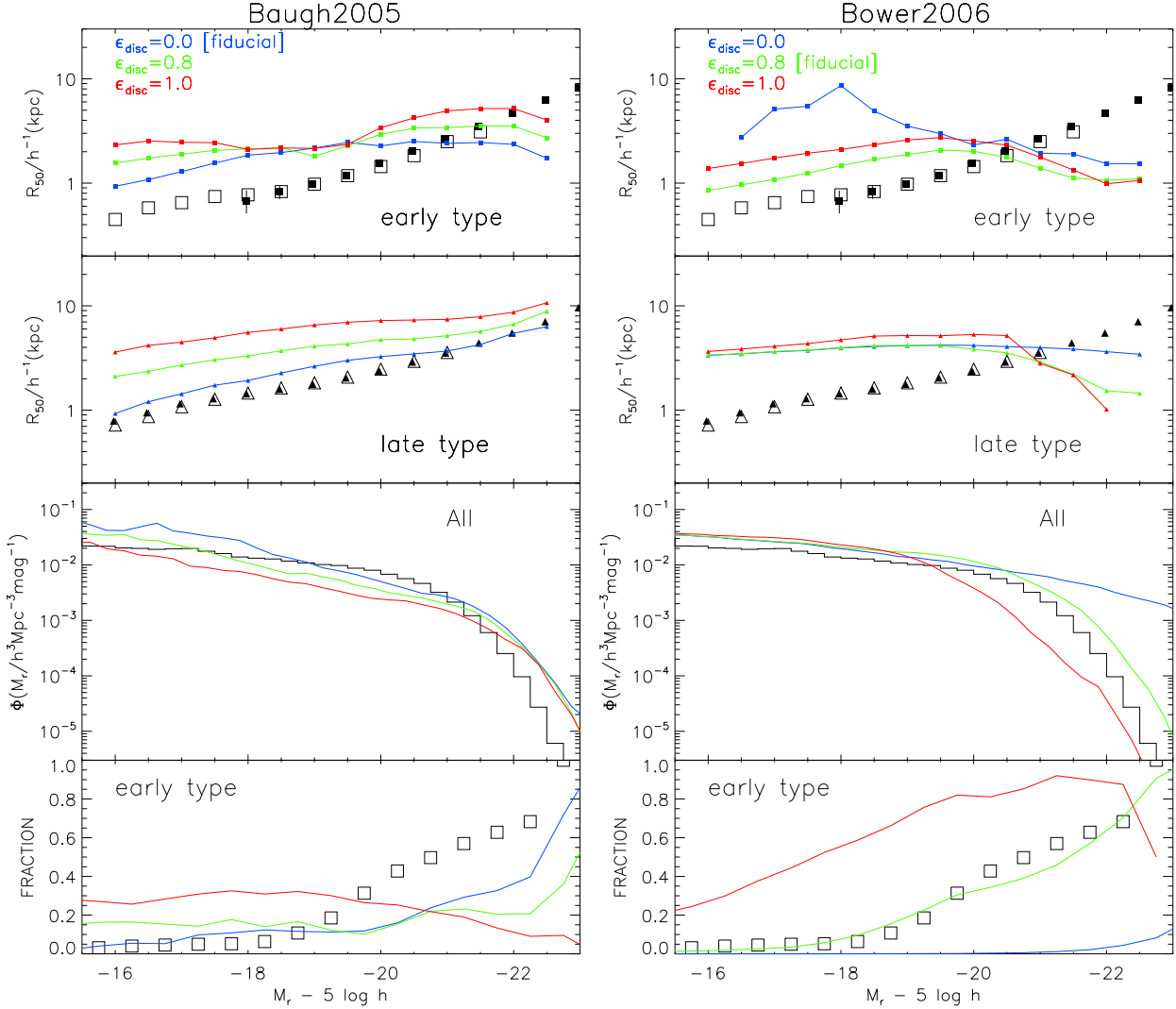}
\caption{
Similar to Fig.~\ref{Shenmedvhot}, but varying the disk instability threshold $\epsilon_{\rm disk}$ in Eq.~\ref{eqinstab}.
}
\label{Shenmedinstab}
\end{figure*}
\begin{figure*}
\includegraphics[width=17cm]{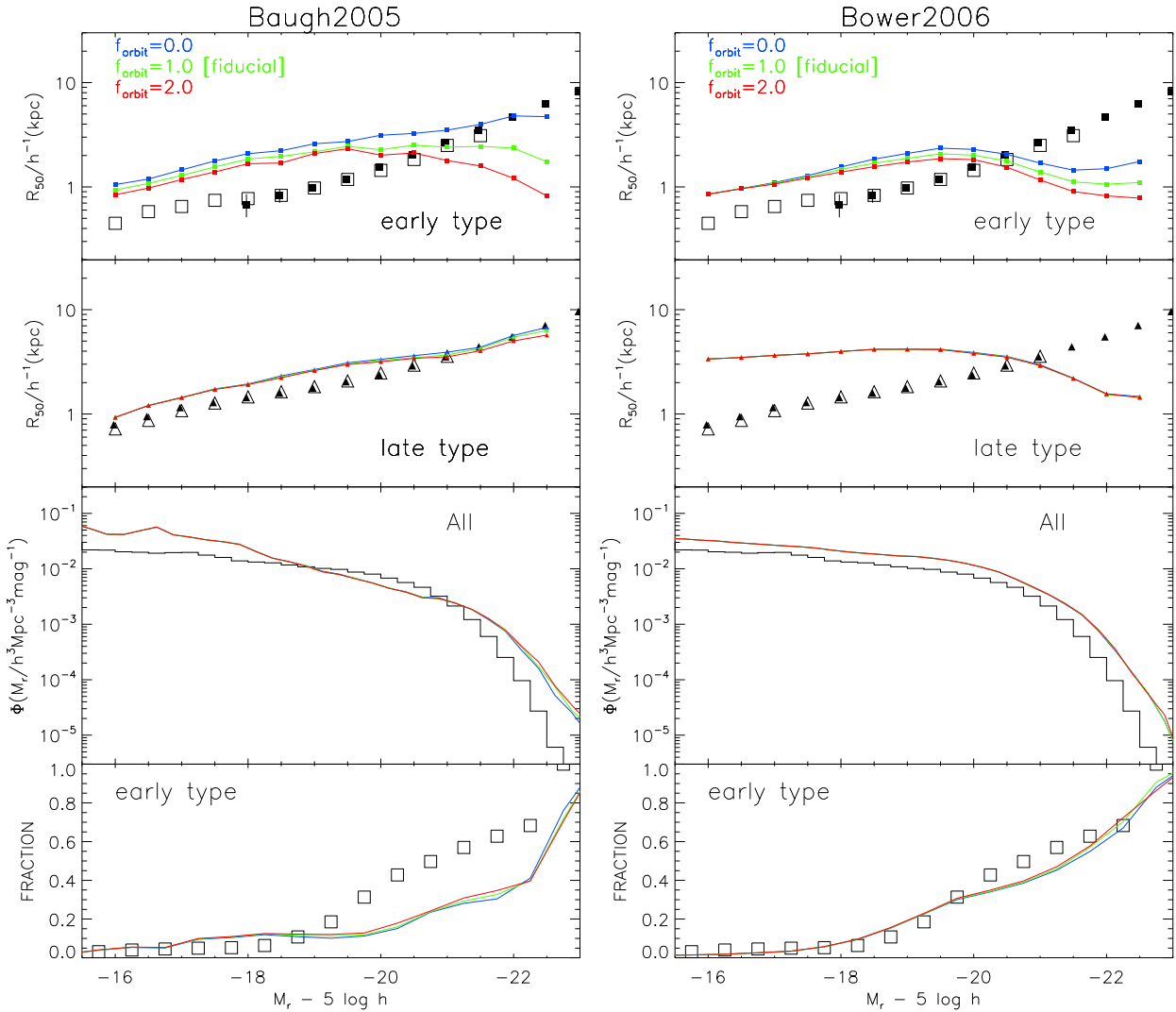}
\caption{
Similar to Fig.~\ref{Shenmedvhot}, but varying the $f_{\rm orbit}$ parameter in Eq.~\ref{rnew}
}
\label{Shenmedforbit}
\end{figure*}

\begin{figure*}
\includegraphics[width=17cm]{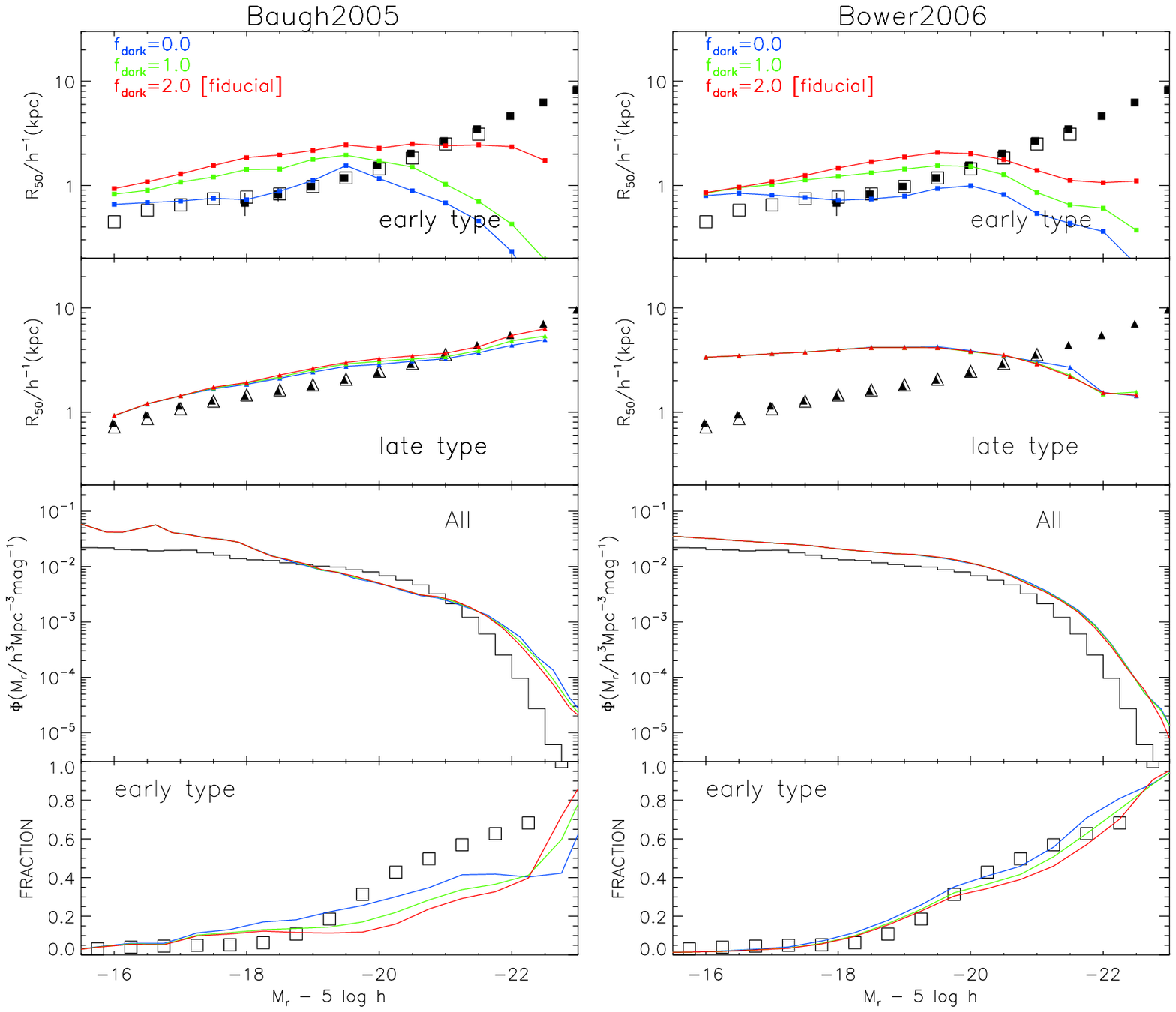}
\caption{
Similar to Fig.~\ref{Shenmedvhot}, but varying the $f_{\rm dark}$ parameter in Eq.~\ref{DMcontrib}
}
\label{ShenmedDMc}
\end{figure*}

Following Sh03, we take $c<2.86$ to define the disk-dominated sample
(recall that pure disk galaxies have $c \sim 2.3$ and pure bulges have
$c \sim 3.3$, as shown by Fig.~\ref{BTvsc}). Besides the selection
introduced by use of the SDSS spectroscopic sample ($r_{\rm
Pet}<17.77$), further selection criteria are required in the size
distribution analysis.  The size measurement for compact galaxies
could be affected by the point-spread function of the image or because
these objects could be misclassified as stars by the SDSS imaging
processing software.  To minimize such contamination, Sh03 selected
galaxies with angular sizes $R_{50} > 1.6''$ (this excludes only a few
percent of the galaxies). Sh03 further restricted the sample to
galaxies with surface brightness $\mu_{50}\leq 23.0$ mag
arcsec$^{-2}$, and apparent magnitude in the range $r_{\rm min}
(\theta,\phi) \leq r \leq r_{\rm max} (\theta,\phi)$ with, typically,
$r_{\rm min} \sim 15.0$ and $r_{\rm max} \sim 17.77$, and redshift $z
\geq 0.005$.

We apply the same criteria as used by Sh03 to the low redshift
NYU-VAGC catalogue. This requires us to recalculate the value of
$V_{\rm max}$ needed to construct statistical distributions, to take
into account the bright magnitude limit, size cut and surface
brightness cut.  Note that although the Sh03 sample is from the
smaller DR1, it contains more galaxies than the NYU-VAGC sample used
here because it extends to higher redshift.

The correlation of size with luminosity for disk-dominated galaxies is
shown in the upper six panels of Fig.~\ref{AllSizesShort}, in which we
plot the distribution of $R_{50}$ in selected magnitude bins in the
$r-$band.  The {\tt GALFORM} predictions are plotted as unshaded
histograms, with the Baugh2005 results in blue and the Bower2006
results in red. The observed distributions are shown by the yellow
filled histograms. Except for the brightest two magnitude bins shown,
the models tend to overpredict the size of disk-dominated galaxies,
particularly in the case of the Bower2006 model. In the Baugh2005
model, the peak of the distribution shifts to larger sizes with
brightening magnitude, reproducing the trend seen in the
observations. On the other hand in the Bower2006 model, there is
little dependence of disk size on luminosity. Both models display a
larger scatter in sizes than is seen in the data. The panels for
early-types are discussed in the next section.

To further quantify the trend of size with luminosity, we calculate
the median value of $R_{50}$ and plot the results in
Fig.~\ref{Shenmedvhot}, where the continuous green line represents the
fiducial {\tt GALFORM} model (the left panel shows the results for the
Baugh2005 model and the right panel for the Bower2006 model) and open
symbols represent the NYU-VAGC data, where we overplot for comparison
(and to check for consistency) the results from Sh03 (filled
triangles).  Fig.~\ref{Shenmedvhot} shows that our analysis of the
size distribution in the NYU-VAGC is consistent with the results of
Sh03. The apparent magnitude cut $r_{\rm min}$ together with using a
low-redshift sample in comparison with Sh03, removes the galaxies
brighter than $M_{r}-5\log h = -21.5$.

The SDSS data show an increase of over one decade in $R_{50}$ across
the luminosity range plotted for disk-dominated galaxies. This
increase is reproduced by the predictions of the Baugh2005 model. The
behaviour of the Bower2006 model is quite different, with an
essentially flat size-luminosity relation to $~L_{*}$, followed by a
decrease in size for brighter galaxies. We investigate the impact of
various processes on the form of the size predictions in Section
4.5.3.

\subsubsection{Bulge-dominated galaxies}

We select a bulge-dominated sample by taking those galaxies with 
concentration index $c>2.86$. In the lower six panels of 
Fig.~\ref{AllSizesShort}, we plot the distribution of sizes 
$R_{50}$ for bulge-dominated galaxies for a selection of magnitude 
bins. In general, the model predicts values of $R_{50}$ larger than 
observed, except for the $M_{r}-5 \log h = -20.5$ bin for Bower2006 
and  $M_{r}-5 \log h = -21.25$ for Baugh2005. As for the case of 
disk-dominated galaxies, the predicted scatter in sizes is larger 
than observed.  

We plot the median size of the bulge-dominated samples in the top row
row of Fig.~\ref{Shenmedvhot}. The predicted size-luminosity relation
is flatter than observed, turning over at the brightest magnitudes
plotted.  The brightest galaxies are three to five times than smaller
than observed, confirming the conclusion reached by \citep{Al07}.

As the DR4 data set we are working with covers a larger solid angle 
than the sample used by Sh03, the combined set of data measurements 
covers a wdier range of magnitudes than can be reached by either 
sample alone. Again where there is overlap, we find that our analysis 
of DR4 is consistent with the results of Sh03.

\subsubsection{Sensitivity of the predictions to physical ingredients}

The calculation of sizes involves several components as outlined in \S2.2. 
Given that this is 
the area in which, overall, the model predictions agree least well with the 
observations, it is instructive to vary some of the physical ingredients 
of the model to see if the agreement can be improved. In the tests which 
follow, we vary the strength of one ingredient at a time and assess the 
impact on the size-luminosity relation. We also show the effect of the 
parameter change on the form of the overall galaxy luminosity function 
and the mix of morphological types. These variant models are not intended 
to be viable or alternative models of galaxy formation, but instead allow 
us to gain some physical insight into how the model works.

\begin{itemize}
\item[(i)]{\it The strength of supernova feedback.}

SN feedback plays an important role in setting the sizes of disk
galaxies, by influencing in which haloes gas can remain in the cold
phase to form stars. Cole et~al. (2000) demonstrated that increasing
the strength of SN feedback results in more gas cooling to form stars
in more massive haloes, which leads to larger disks. Conversely,
reducing the feedback allows gas to cool and form stars in smaller
haloes resulting in smaller discs. The strength of SN feedback is
parameterized using $V_{\rm hot}$ and $\alpha_{\rm hot}$ as shown in
Eq.~\ref{eqsnfeedback}. The adopted values for these parameters are:
$V_{\rm hot} = 300\,{\rm km\,s}^{-1}$ and $\alpha_{\rm hot} = 2$ in
the Baugh2005 model and $V_{\rm hot} = 485\,{\rm km\,s}^{-1}$ and
$\alpha_{\rm hot} = 3.2$ in the Bower2006 model. We perturb the models
by increasing and reducing the value of $V_{\rm hot}$ to its double
and half the fiducial value in each model, and plot the results in
Fig.~\ref{Shenmedvhot}. The normalization of the size-luminosity
relation for disk-dominated galaxies changes as expected on changing
the strength of supernova feedback, moving to larger sizes on
increasing $V_{\rm hot}$ and smaller sizes on reducing $V_{\rm hot}$.
Reducing $V_{\rm hot}$ in the Baugh2005 model leads to better
agreement with the observed size-luminosity relation for
disk-dominated galaxies, at the expense of producing slightly more
faint galaxies. Similar trends are seen in the predictions for the
size-luminosity relation of bulge-dominated galaxies. Note that there
are very few bulge-dominated galaxies at faint magnitudes in the
Bower2006 model, hence the noisy size-luminosity relation in this
region. Changing the strength of feedback in this way has little
impact on the slope of the size-luminosity relation.

\item[(ii)] {\it The threshold for disks to become unstable.}

The threshold for a disk to become unstable is set by the parameter
$\epsilon_{\rm disk}$ (see Eq.~\ref{eqinstab}). We show the result of
varying this threshold in Fig.~\ref{Shenmedinstab}. In the case of the
Bower2006 model, we increase and reduce the threshold from the
fiducial value of $\epsilon_{\rm disk}=0.8$; increasing the threshold
means that more disks become unstable. The original Baugh2005 model
does not test for the stability of disks, so in this case we switch
the effect on and try two different values for the threshold. The
result of turning on dynamical instabilities is straightforward to
understand in this model.  For a given mass and rotation speed of
disk, the stability criteria $\epsilon \propto \sqrt{r_{\rm disk}}$,
and so disks with smaller values of $r_{\rm disk}$ will preferentially
be unstable. The removal of small disks raises the median disk size
but reduces the fraction of galaxies that are disk-dominated. 
The impact of varying the stability threshold on the
Bower2006 model is less easy to interpret: even though the fraction of
faint disk-dominated galaxies fails dramatically on increasing the
threshold, there is little change in the median size of the surviving
disks. This change has a bigger impact on the size of bulge-dominated
galaxies, with a sequence that is inverted compared with the Baugh2005
model. One result that is easily understood is the response of the
luminosity function.  The burst of star-formation which can accompany
the transformation of a dynamically unstable disk into a bulge is an
important channel for generating black hole mass in the Bower2006, as
discussed in Section 2.2.  If fewer disks become unstable, less mass
is converted into black holes and AGN feedback has less impact on the
cooling flows in massive haloes, leading to too many bright galaxies.

\item[(iii)] {\it The orbital energy of merging galaxies}

The parameter $f_{\rm orbit}$ quantifies the orbital energy of two galaxies 
which are about to merge (Eq.~\ref{Eorbit}). Our standard choice in both 
models is $f_{\rm orbit}=1$, in which case Eq.~\ref{Eorbit} is equal to the 
energy of two point masses in a circular orbit at a separation of 
$r_{1}+r_{2}$. We vary the value of $f_{\rm orbit}$ trying 
$f_{\rm orbit} = 0$, which corresponds to a parabolic orbit, and 
$f_{\rm orbit}=2$. The results are plotted in Fig.~\ref{Shenmedforbit}. 
As expected, the median size of disk-dominated galaxies is unaffected by 
varying $f_{\rm orbit}$. Increasing the value of $f_{\rm orbit}$ makes 
bulge-dominated galaxies smaller, with a larger effect seen for brighter 
galaxies. A smaller value of $f_{\rm orbit}$ improves the shape of the 
size-luminosity relation of bulge-dominated galaxies in the Baugh2005 
model; however, faint bulge-dominated galaxies are still too large after 
making this change.  

\item [(iv)]{\it The contribution of the dark matter in galaxy mergers} 

The parameter $f_{\rm dark}$ controls the amount of dark matter
associated with model galaxies during merger evenst (see
Eq.~\ref{DMcontrib}), which has an impact on the size of the merger
remnant, through Eq.~\ref{Eorbit} and Eq.~\ref{rnew}.  We run the
models using values of $f_{\rm dark}=1$ and $f_{\rm dark}=0$, the
latter of which corresponds to the case of a galaxy without
participating dark matter.  We can see that the reduction of $f_{\rm
dark}$ from the fiducial value leads to smaller sizes for the early
type galaxies in both models.  The effect is particularly important at
bright magnitudes. As expected, there is no change in the predicted
sizes for late type galaxies. We do not find, either, a variation in
the luminosity function, but there is an increase in the fraction of
early type galaxies, particularly at intermediate magnitudes. We can
see that we can improve the sizes of early type galaxies, matching
with those inferred from SDSS observations for galaxies with
magnitudes fainter than $L\star$. However, this change is
counterproductive at bright magnitudes, resulting in even smaller
sizes.

\end{itemize}

\subsubsection{What drives the slope of the size-luminosity relation?}

\begin{figure}
\includegraphics[width=8.5cm]{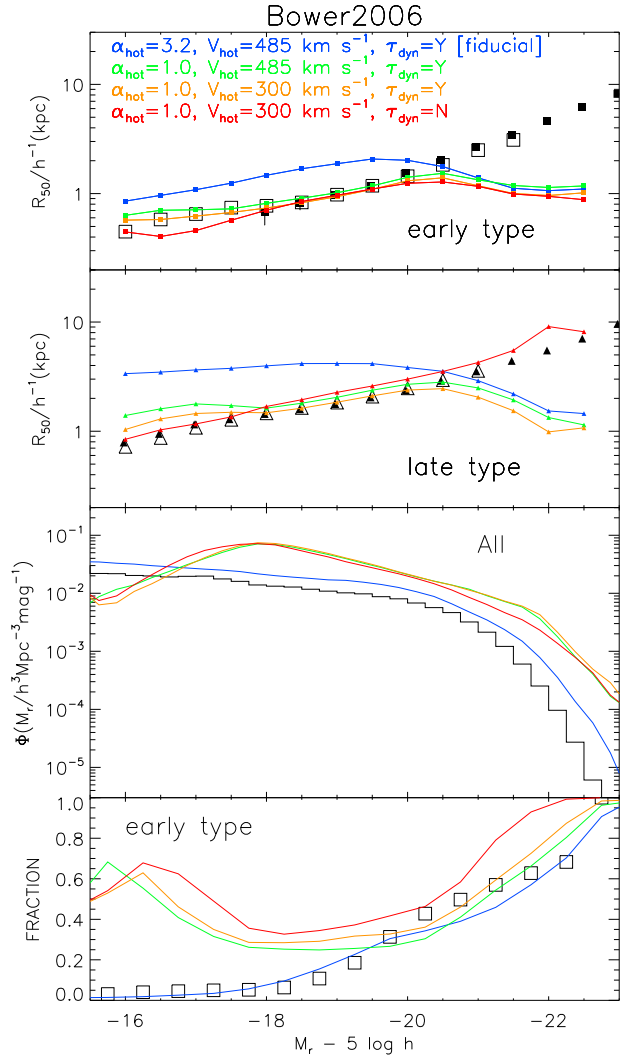}
\caption{ The impact of cumulative parameter changes for a sequence of
  models, starting from the fiducial model of Bower2006 (blue). One
  parameter is changed in each step, but unlike the cases presented in
  Figs.~16, 17, 18, 19, the change made is relative to the previous
  model in the sequence.  The key lists the parameter change relative
  to the previous model in the sequence.  The upper two panels show
  the size-luminosity relation for bulge and disk-dominated galaxies
  respectively. The third panel shows the luminosity function and the
  bottom panel shows the morphological mix of galaxies as a function
  of luminosity.  }
\label{flattosteep}
\end{figure}

We have seen that the prediction of the Bower2006 model for the
size-luminosity relation of disk-dominated galaxies is much flatter
than that of the Baugh2005 model (Fig.~\ref{Shenmedvhot}). Moreover
the Baugh2005 model is in better agreement with the observed relation.
In the previous section we varied selected model parameters one at a
time relative to the fiducial model, to show their impact on the model
size-luminosity relation.  In this exercise, the most dramatic change
in the Bower2006 predictions resulted from varying the strength of SNe
feedback.  Reducing the value of the parameter $V_{\rm hot}$, which
sets the ``pivot'' velocity below which SNe feedback has a strong
impact, leads to a shift in the size-luminosity relation to smaller
sizes, with an improved match to the observed relation recovered for
intermediate luminosity galaxies. In this section, we investigate the
effect of varying several parameters together, essentially moving from
the Bower2006 parameters for SNe feedback and the star formation
timescale in disks, towards a set which more closely resembles that
used in the Baugh2005 model.  The size-luminosity relations for disk-
and bulge-dominated galaxies are plotted in Fig.~\ref{flattosteep} for
a sequence of models.  The starting point is the fiducial Bower2006
model. For each step in the sequence, one parameter is varied relative
to the previous step, as shown in the key in
Fig.~\ref{flattosteep}. The first change made is to the value of
$\alpha_{\rm hot}$, which controls the slope of the SNe
feedback. Changing from the Bower2006 value of $\alpha_{\rm hot}=3.2$
to $\alpha_{\rm hot}=1$ gives a much improved match to the observed
size-luminosity relation, particularly for intermediate
luminosities. Faint disk-dominated galaxies are still somewhat too
large, and bright galaxies in general are still too small. The next
step is to retain the above change to $\alpha_{\rm hot}$, and also to
change the value of $V_{\rm hot}$ to that used in Baugh2005.  This
results in a modest improvement in the size-luminosity relation for
faint galaxies. Finally, the scaling of the quiescent star formation
timescale with the disk dynamical time is switched off. The resulting
size-luminosity relation is now in very good agreement with the
observations for disk-dominated galaxies. However, bright
bulge-dominated galaxies are still too small. Furthermore, the
luminosity function and the predicted fraction of early types with
luminosity are now much poorer matches to observations than in the
fiducial Bower2006 model (lower two panels of Fig.~\ref{flattosteep}).

In summary, even though the sizes of disk-dominated galaxies can be
brought into reasonable agreement with observations in a variant of
the Bower2006 model with modified SNe feedback and disk star formation
timescale, this is at the expense of agreement with the observed
luminosity function and early-type fraction. Furthermore, neither the
fiducial Baugh2005 model nor the fiducial or modified Bower2006 models
are able to reproduce the observed sizes of early-type galaxies, in
particular for bright galaxies. No single parameter change seemed able
to solve the latter problem.  The most promising area to explore
further appears to be the modelling of galaxy mergers; changing the
amount of orbital energy brought in by merging galaxies in our
prescription led to an increase in the sizes of the brightest
bulge-dominated galaxies.

\section{SUMMARY AND DISCUSSION}

Observations of local galaxies have always played a central role in
setting the parameters of galaxy formation models. However, the huge
size of the SDSS sample combined with the quality and uniformity of
the data allow much more precise and exacting tests of the physics of
such models than was previously possible. To take full advantage of
this opportunity, it is necessary for the model to be able to generate
predictions which are as close as possible to the measurements made
for real galaxies.  In this paper we use the {\tt GALFORM} model,
which predicts the size of both the disk and bulge components of
galaxies. Hence we are able to take the model output for the
luminosity and scale-lengths of a galaxy's disk and bulge and turn
these into predictions for the quantities measured for SDSS galaxies:
the Petrosian magnitude, the radius containing 50\% of the Petrosian
flux ($R_{50}$), the concentration parameter ($c$) and the S\'{e}rsic
index ($n$), the latter two quantities providing descriptions of the
light profile of the galaxy.

The first major result of this work is to understand the correlation
between different indicators of galaxy morphology. The concentration
parameter, S\'{e}rsic index and bulge-to-total ($B/T$) luminosity ratio
have all been used to divide galaxies into morphological classes
\citep[e.g][]{Bershady00,Hog04,Ben07}.  The $B/T$ ratio is easy to
compute theoretically, yet is perhaps the hardest of these quantities
to measure observationally. Both the $c$-$B/T$ and $n$-$B/T$ planes
show scatter. This can be traced to the ratio of the disk and bulge
scale-lengths; galaxies with different values of this ratio occupy
different loci in the $c$-$B/T$ and $n$-$B/T$ planes. The scatter is
particularly large in the case of the S\'{e}rsic index $n$ versus $B/T$.
The scatter would be even larger if we simply generated galaxies by
hand, taking the ratio of disk and bulge scale-lengths from a
grid. The scatter we find is limited by the distribution of $r_{\rm
d}/r_{\rm b}$ values predicted by {\tt GALFORM}.

We compared the predictions of two different versions of the {\tt
GALFORM} model with SDSS data: that of \citet{Bau05}, which has a
top-heavy IMF in starbursts and feedback from superwinds, and
\citet{Bow06}, which has AGN feedback and a normal IMF in all modes of
star formation. In the first stage of the comparison, none of the
model parameters were adjusted to improve the fit to the data. The
models gave reasonable matches to the total galaxy luminosity
function, with the \citeauthor{Bow06} model giving the best overall
match to the shape. The match to the luminosity function of different
colour subsamples is less impressive; both models overpredict the
number of bright blue galaxies and fail to match the number of red
galaxies.  The \citeauthor{Bow06} model has a strongly bimodal colour
distribution, whereas the \citeauthor{Bau05} model shows only weak
bimodality. The \citeauthor{Bow06} model agrees better overall with
the observed colours in SDSS, although the predicted bimodality
appears somewhat too strong, and the positions of the peaks in the
colour distribution do not agree exactly with what is observed.

Another clear success of the models is in predicting the correct trend
of morphological type with galaxy luminosity. We used all three
morphological indicators (concentration parameter, S\'{e}rsic index and
bulge-to-total luminosity ratio from disk+bulge fits) to separate
galaxies into disk-dominated and bulge-dominated types. In the SDSS
data, the fractions of these types are found to shift from being
almost completely disk-dominated at low luminosity to almost entirely
bulge-dominated at high luminosity, though with differences in the
detailed behaviour depending on which morphological indicator is
used. Both the \citet{Bau05} and \citet{Bow06} models successfully
reproduce this general trend, although the \citeauthor{Bow06} model
agrees better in detail with the observed behaviour at intermediate
luminosities. Both models qualitatively reproduce the observed
correlation of colour with morphology (with bulge-dominated galaxies
on average being redder than disk-dominated galaxies at every
luminosity), although quantitatively the \citeauthor{Bow06} model
agrees better with the SDSS data, the \citeauthor{Bau05} model
predicting too many red disk-dominated galaxies.

Perhaps the most serious shortcoming of the models is in the predicted
galaxy sizes. Whilst the \citeauthor{Bau05} model gives a very good
match to the luminosity-size relation for disks, the sizes of
bulge-dominated galaxies do not match the observations. The slope of
the size-luminosity relation for bulge-dominated galaxies in the Baugh
et~al. model matches the observations for faint galaxies, but the
normalization is too high. Brighter than $L_{*}$, the predicted
relation flattens, with the consequence that the brightest
bulge-dominated galaxies are around a factor of three too small (see
also \citealt{Al07}). The situation is worse for the sizes predicted
by the \citeauthor{Bow06} model; in this case the size - luminosity
relation is flat for disk-dominated galaxies, while for
bulge-dominated galaxies the predicted sizes at high luminosities fall
even further below the observed relation than for the
\citeauthor{Bau05} model. We have demonstrated that a steeper slope
for the size-luminosity relation for disk-dominated galaxies can be
recovered in the \citeauthor{Bow06} model if we set some physical
processes to have the same parameters as used in the
\citeauthor{Bau05} model. The primary improvement in the model
predictions is seen on reducing the strength of SNe feedback.  Also,
by adjusting the star formation timescale in disks by switching off
the dependence on the disk dynamical time, we can recover the observed
slope of the size-luminosity relation even at high
luminosities. However, this improvement in the size - luminosity
relation comes at the expense of producing too many galaxies overall.

The differences between the predictions of the two models for the
sizes of disk-dominated galaxies lie in the revised cooling model
adopted by \citet{Bow06}, the strength of supernova feedback, the
inclusion of AGN feedback and the inclusion of dynamical instabilities
for disks. In the \citeauthor{Bow06} model, gas which is reheated by
supernova feedback is reincorporated into the hot halo on a shorter
timescale than in the \citet{Bau05} model.  Neither model is able to
match the observed size of bright bulge-dominated galaxies. We explore
a range of processes in the models, varying the strength of supernova
feedback, changing the threshold for disks to become unstable, and
changing the prescription for computing the size of the stellar
spheroid formed in a galaxy merger.  The latter seems the most
promising solution, at least in the case of the \citeauthor{Bau05}
model. If we neglect the orbital energy of the galaxies which are
about to merge (i.e. setting the parameter $f_{\rm orbit}=0$), then
the sizes of bright galaxies are in much better agreement with the
observed sizes (though the faint bulge-dominated galaxies are still
too large). A similar change in the predictions results from ignoring
the adiabatic contraction of the dark matter halo in response to the
gravity of the disk and bulge \citep{Al07}.

The {\tt GALFORM} model is one of the few able to make the range of
predictions considered in this paper and hence to take full advantage
of the constraining power of the SDSS. The model for disk sizes works
well under certain conditions, albeit with too much scatter.  Our
analysis suggests that the problems with disk sizes and colours are
connected to the treatment of gas cooling and supernova feedback,
while the problems with spheroid sizes are probably due to an
overly-simplified treatment of the sizes of galaxy merger remnants.
The prescription used to compute the size of spheroids is in need of
improvement, which will require the results of numerical simulations
of galaxy mergers.  This study highlights the need to make careful and
detailed comparisons with observational data in order guide
improvements in the treatment of physical processes in galaxy
formation models.

\section*{Acknowledgements} 
We thank Shiyin Shen for kindly providing results in electronic form
to include in our figures and the anonymous referee for providing a
useful report.  This work was supported in part by the European
Commission's ALFA-II project through its funding of the Latin American
European Network for Astrophysics and Cosmology (LENAC), by a Science
and Technology Facilities Council rolling grant and by the Royal
Society. JEG acknowledges receipt of a fellowship funded by the
European Commission's Framework Programme 6, through the Marie Curie
Early Stage Training project MEST-CT-2005-021074. AJB acknowledges the
support of the Gordon and Betty Moore Foundation.

Funding for the creation and distribution of the SDSS Archive has been
provided by the Alfred P. Sloan Foundation, the Participating
Institutions, the National Aeronautics and Space Administration, the
National Science Foundation, the U.S. Department of Energy, the
Japanese Monbukagakusho and the Max Planck Society. The SDSS Web site
is http://www.sdss.org/.  The SDSS is managed by the Astrophysical
Research Consortium (ARC) for the Participating Institutions. The
Participating Institutions are The University of Chicago, Fermilab,
the Institute for Advanced Study, the Japan Participation Group, The
Johns Hopkins University, Los Alamos National Laboratory, the
Max-Planck-Institute for Astronomy (MPIA), the Max-Planck-Institute
for Astrophysics (MPA), New Mexico State University, Princeton
University, the United States Naval Observatory, and the University of
Washington.

\appendix
\section[]{Changes in the photometry of SDSS galaxies from DR1 to DR4}

\begin{figure}
\includegraphics[width=8.7cm]{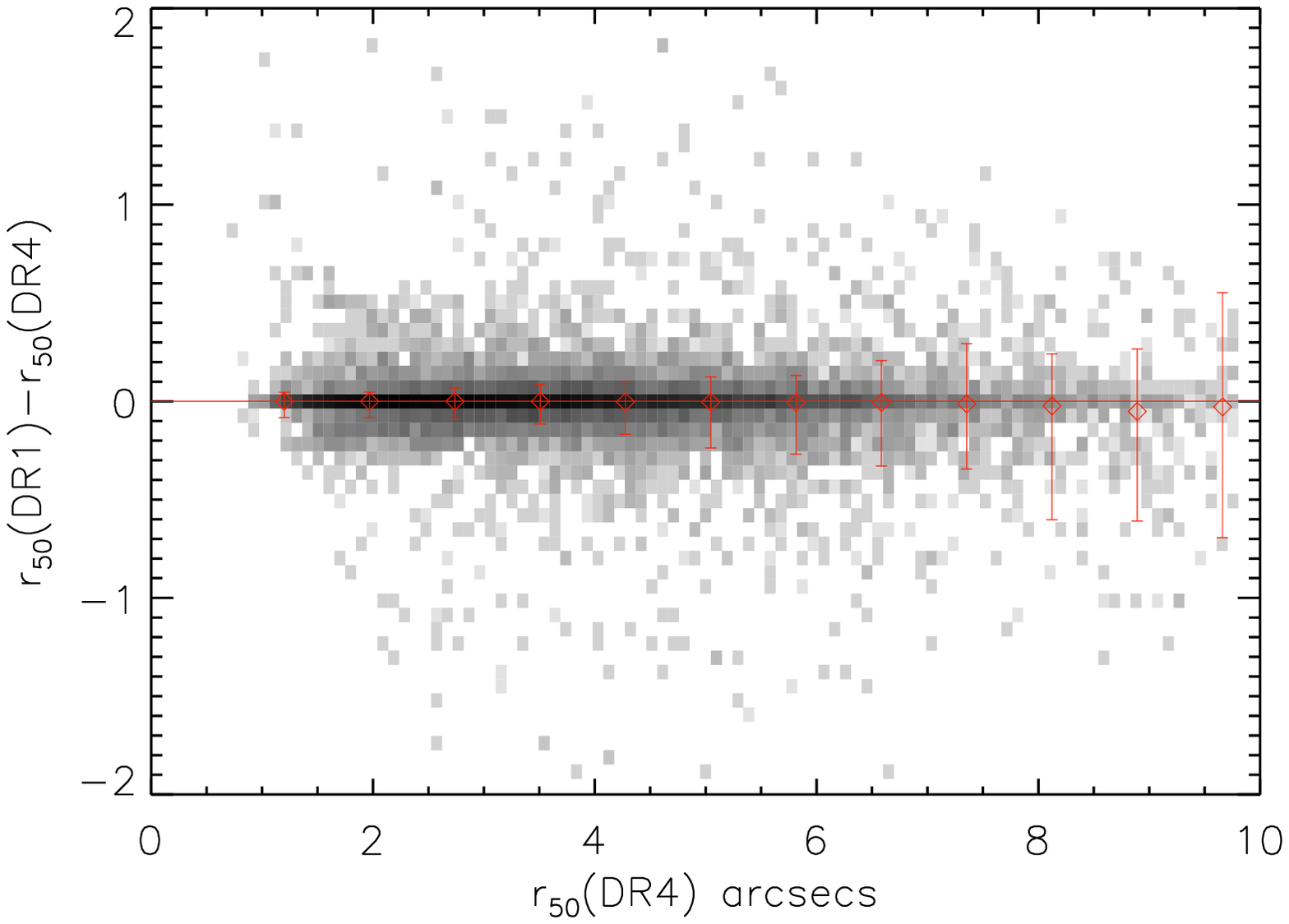}
\caption{ The difference in angular radius (in arcsec) enclosing 50\%
of the Petrosian magnitude, $r_{50}$, between for the same galaxies
identified in DR1 and DR4. The shading reflects the logarithm of the
density of galaxies. The points show the median difference in size and
the bars show the 10-90 percentile range of this distribution.  }
\label{r50dr1dr4}
\end{figure}

\begin{figure}
\includegraphics[width=8.7cm]{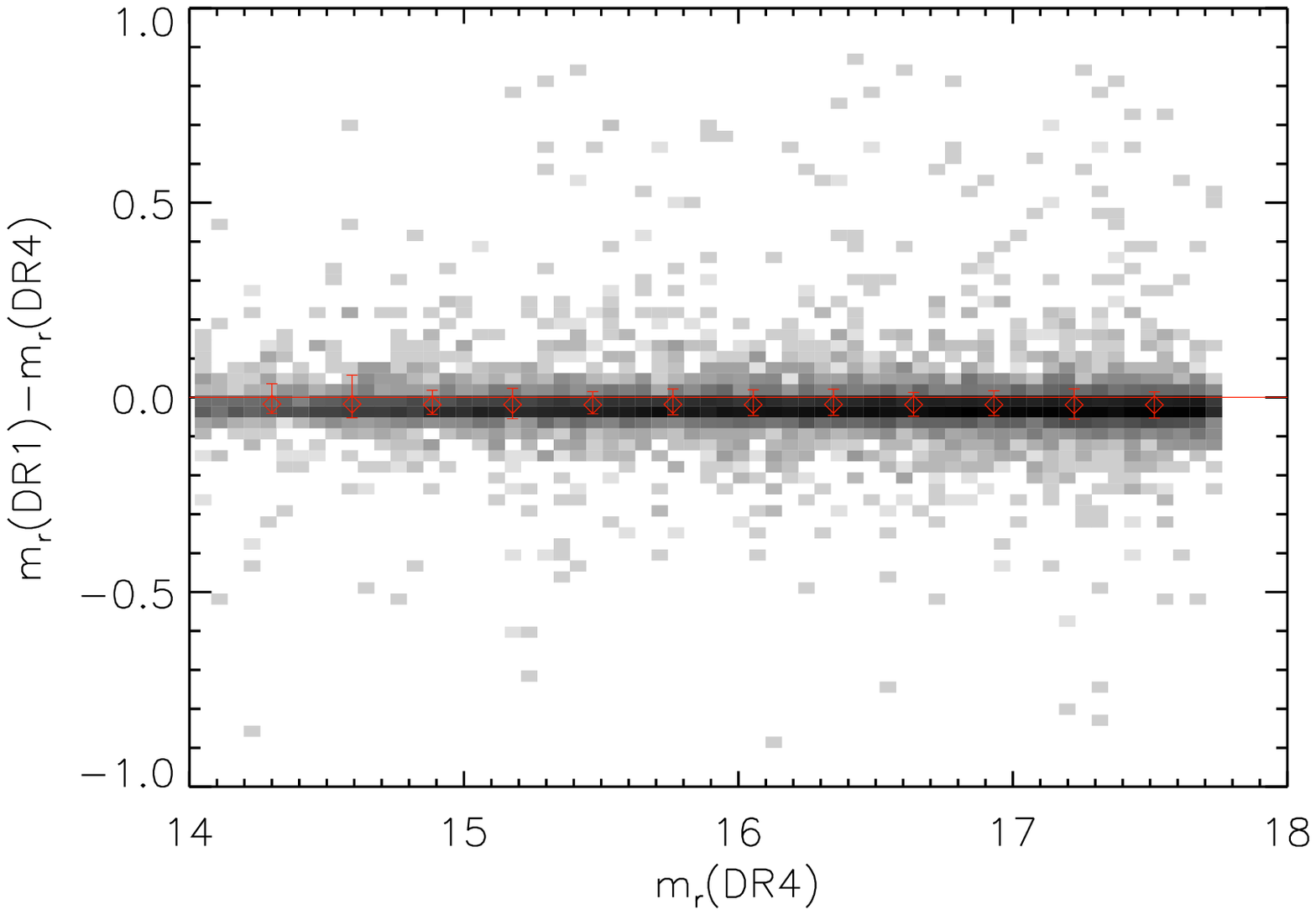}
\caption{ The difference in the Petrosian magnitude recorded in DR1
and DR4 for a matched sample of galaxies. The shading and symbols have
the same meaning as in Fig.~\ref{r50dr1dr4}.  }
\label{rappdr1dr}
\end{figure}

\begin{figure}
\includegraphics[width=8.7cm]{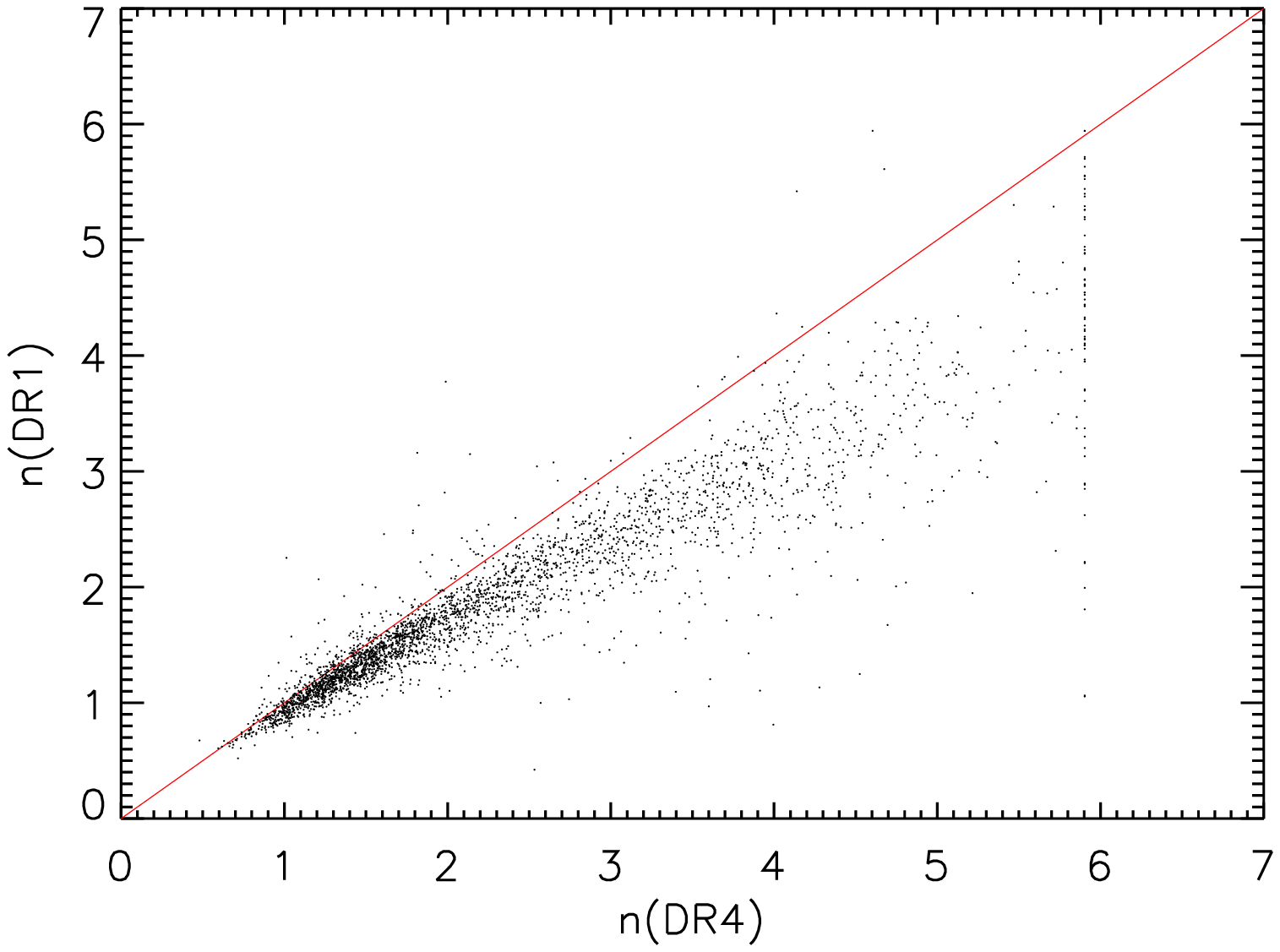}
\caption{D
The S\'{e}rsic index in DR1 plotted against that measured for the same 
galaxy in DR4. The shading and symbols have the same meaning 
as in Fig.~\ref{r50dr1dr4}.
}
\label{ndr1dr}
\end{figure}

The photometric and spectroscopic pipelines for processing SDSS data
have been refined on subsequent data releases, particularly between
DR1 and DR2. In DR2, all the SDSS data was re-analyzed to apply
improvements to the processing of images (magnitude modelling, image
deblending) and spectra (extraction of radial velocities,
spectrophotometry). It is instructive to test whether any of the
photometric properties used in this paper to constrain the models have
changed appreciably between data releases. Uncertainty in the
extraction of properties from observational data puts a limit on how
well we can expect the models to reproduce the observational
results. In this appendix, we compare galaxy sizes, Petrosian
magnitudes and S\'{e}rsic index values between DR1 and DR4.

To perform the comparison between measurements in different data
releases, we need to be sure that we are looking at the same galaxy in
each version of the catalogue. This is not a trivial exercise, since
revisions to the algorithm used to deblend close or merged images mean
that a single object in DR1 could appear as multiple objects in
DR4. The match is made by requiring that a DR4 galaxy should be closer
than $1.2 \arcsec$ on the sky (which is equivalent to 3 SDSS pixels,
each of $0.396 \arcsec$). This is close to the smallest angular size
found for galaxies used in the comparison. With this criteria, we were
able to find DR4 counterparts for 95\% of the galaxies from DR1.

We first look at the difference in the value of angular radius
enclosing 50\% of the Petrosian light, which is plotted in
Fig.~\ref{r50dr1dr4} for galaxies with $z<0.05$. Here we plot the
logarithm of the number density of galaxies in greyscale to expand the
dynamic range of the shading. The points with error bars show the
median difference in size between the two data releases, with the bars
showing the 10-90 percentile range of the distribution. Although there
is scatter in the sizes between data releases, there is no evidence
for any systematic differences.

Next we repeat this comparison for the Petrosian magnitude, which is
shown in Fig.~\ref{rappdr1dr} for galaxies with $z<0.05$. In this case
there is a small systematic effect, with the median shift being ~-0.04
mag between DR1 and DR4 i.e.  a galaxy is typically brighter in DR1
and it appears in DR4.

Finally we compare the S\'{e}rsic index between DR1 and DR4. For DR1, 
we use the S\'{e}rsic index calculated by \citet{Bla03}. The analysis presented 
by these authors corresponds to a larger area than DR1, but relies on 
a pre-DR1 version of the photometric analysis software. In a subsequent 
publication, \citep{Bl05}, the algorithm used to compute the S\'{e}rsic 
index was updated in order to account for a bias in the results. 
\citep{Bl05} demonstrated this effect by feeding a pure bulge with $n=4$ 
into the algorithm. With the original method, a S\'{e}rsic index of $n=2.7$ 
was recovered. Using the improved algorithm, the result was increased to 
$n=3.5$, a much smaller bias. The comparison between the S\'{e}rsic index in 
DR1 and DR4 is plotted in Fig.~\ref{ndr1dr}.The DR4 S\'{e}rsic index is 
generally larger than in DR1, particularly for $n>1$. The revised algorithm 
sometimes fails to find a suitable value for $n$, in which case $n=6$ 
was assigned. This comparison shows the difficulty in extracting the value 
of the S\'{e}rsic index for galaxies, and gives an indication of how closely we 
should expect the models to agree with the observational results.

\section[]{The correlation between S\'{e}rsic index and bulge to total 
luminosity ratio.}

\begin{figure}
\includegraphics[width=8.7cm]{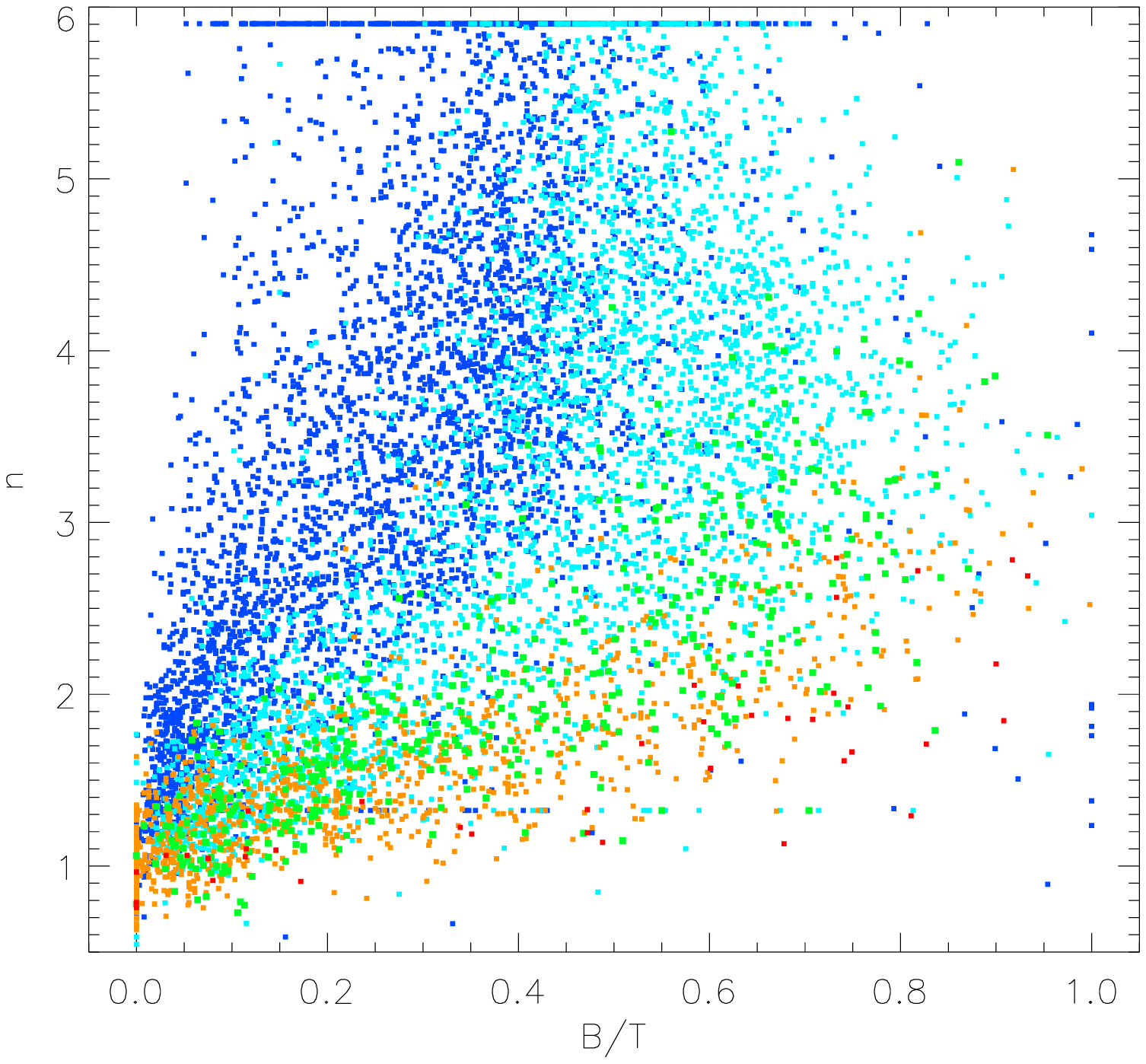}
\caption{ The S\'{e}rsic index extracted from the NYU-VAGC plotted
against the bulge-to-total luminosity ratio (B/T) as determined by
\citet{Ben07}. The colour coding reflects the ratio of the disk and
bulge radii, which blue indicating a ratio of $r_{\rm d}/r_{\rm b}=4$
and red indicating $r_{\rm d}/r_{\rm b}=0.25$, as in Fig. 3.  }
\label{ALLBTvsN}
\end{figure}

In Section 3, we compared different indicators of galaxy morphology,
the concentration, S\'{e}rsic index and bulge-to-total luminosity ratio
using the output of {\tt GALFORM}. We found considerable scatter
between these quantities, particularly in the S\'{e}rsic index -
bulge-to-total ratio plane. This is driven by the ratio of the disk
and bulge scale-lengths; galaxies with a given ratio of scale-lengths
occupy a particular locus in the plane.

We can now repeat this comparison for SDSS galaxies. \citet{Ben07}
calculated the disk and bulge radii ($r_{d}$ and $r_{b}$) and
bulge-to-total luminosity ratio, $B/T$, for a sample of galaxies from
the SDSS EDR.  In Fig.~\ref{ALLBTvsN} we plot the raw uncorrected
values of $B/T$ found by \citet{Ben07} against the S\'{e}rsic index $n$
for these galaxies given by the NYU-VAGC used in this paper. The
galaxies are colour-coded in the same way as for the {\tt GALFORM}
sample plotted in Fig.~\ref{B05BTvsN}; the largest ratio of disk to
bulge radii is shown by blue points and the smallest ratio by red
points. This plot looks qualitatively similar to the one obtained in
Section 3 using {\tt GALFORM} output, but with much more scatter.  As
found by \citeauthor{Ben07}, there is a deficit of galaxies with high
$B/T$.

\bsp

\label{lastpage}

\end{document}